\pdfoutput=1
\documentclass[a4paper,11pt]{article}
\usepackage[margin=1in]{geometry}
\usepackage[style=numeric,sorting=none]{biblatex}
\usepackage{amsmath,amssymb,amsfonts}
\usepackage{algorithmic}
\usepackage{graphicx,color,xcolor}
\usepackage{textcomp}
\usepackage{epstopdf}
\def\BibTeX{{\rm B\kern-.05em{\sc i\kern-.025em b}\kern-.08em
    T\kern-.1667em\lower.7ex\hbox{E}\kern-.125emX}}
\usepackage{microtype}
\usepackage{graphicx}
\usepackage{subcaption}
\usepackage{booktabs} 

\usepackage{framed}
\newcommand{\ra}[1]{\renewcommand{\arraystretch}{#1}}
\usepackage{tablefootnote}

\usepackage{hyperref}
\usepackage{comment,cleveref}
\usepackage{bm}
\usepackage{pifont}
\newcommand{\cmark}{\ding{51}}%
\newcommand{\xmark}{\ding{55}}%

\usepackage{amsmath,amsthm,amssymb,amsfonts}

\usepackage{algorithm}
\usepackage{algorithmic}
\usepackage{enumerate}
\usepackage{bbm}
\usepackage[keeplastbox]{flushend}

\theoremstyle{plain}
\newtheorem{theorem}{Theorem}[]
\newtheorem{corollary}{Corollary}[theorem]
\newtheorem{lemma}[]{Lemma}
\newtheorem{proposition}{Proposition}

\newtheorem{definition}{Definition}[]
\newtheorem{assumption}{Assumption}[]

\newtheorem{remark}{Remark}[]

\def\x{{\mathbf x}}
\def\z{{\mathbf z}}
\def\a{{\mathbf a}}
\def\b{{\mathbf b}}
\def\f{{\mathbf f}}
\def\g{{\mathbf g}}
\def\q{{\mathbf q}}
\def\u{{\mathbf u}}
\def\v{{\mathbf v}}

\def\G{{\mathbf G}}

\def\u{{\mathbf u}}
\def\y{{\mathbf y}}

\def\R{{\mathbb{R}}}

\def\G{{\mathcal G}}
\def\E{{\mathcal E}}
\def\F{{\mathcal F}}
\def\O{{\mathcal O}}

\newcommand{\ts}{\textsuperscript}

\begin{document}
\title{Communication-Efficient Variance-Reduced Decentralized Stochastic Optimization over Time-Varying Directed Graphs}
\date{}
\author{Yiyue Chen, Abolfazl Hashemi, Haris Vikalo\thanks{Yiyue Chen and Haris Vikalo  are with the Department of Electrical and Computer Engineering, University of Texas at Austin, Austin, TX 78712 USA. Abolfazl Hashemi is  with the School of Electrical and Computer Engineering, Purdue University, West Lafayette, IN 47907, USA. A preliminary version of this article is presented at the 2021 International Conference on Acoustics, Speech, and Signal Processing (ICASSP) \cite{chen2020communication}.}}

\maketitle

\begin{abstract}
We consider the problem of decentralized optimization over time-varying directed networks. The network nodes can access only their local objectives, and aim to collaboratively minimize a global function by exchanging messages with their neighbors.
Leveraging sparsification, gradient tracking and variance-reduction, we propose a novel communication-efficient decentralized optimization scheme that is suitable for resource-constrained time-varying directed networks. We prove that in the case of smooth and strongly-convex objective functions, the proposed scheme achieves an accelerated linear convergence rate. To our knowledge, this is the first decentralized optimization framework for time-varying directed networks that achieves such a convergence rate and applies to settings requiring sparsified communication. Experimental results on both synthetic and real datasets verify the theoretical results and demonstrate efficacy of the proposed scheme. 
\end{abstract}

\section{Introduction}\label{sec:intro}
Decentralized optimization problems are encountered in a number of settings in control, signal processing, and machine learning \cite{ren2005consensus,jadbabaie2003coordination,nedic2009distributed}. Formally, the goal of a decentralized optimization task is to minimize global objective in the form of a finite sum
\begin{equation}\label{eq:problem}
\min_{\mathbf{x} \in \mathcal{X}} \left[f(\mathbf{x}):=\frac{1}{n}\sum_{i=1}^n f_i(\mathbf{x})\right],
\end{equation}
where $f_i(\mathbf{x}) = \frac{1}{m_i}\sum_{j=1}^{m_i}f_{i, j}(\mathbf{x}): \mathbb{R}^d \to \mathbb{R}$ for $i \in [n]:=\left\{1, ..., n \right\}$ denotes the local objective function that averages loss of $m_i$ data points at node $i$, and $\mathcal{X}$ denotes a convex compact constraint set. The $n$ nodes of the network exchange messages to collaboratively solve (\ref{eq:problem}). Since the communication links between nodes in real-world networks are often uni-directional and dynamic (i.e., time-varying), we model the network by a sequence of directed graphs $\G(t)=(|n|, \E(t))$, where the existence of an edge $e_{i, j} \in \E(t)$ implies that node $i$ can send messages to node $j$ at time step $t$. 

As the networks and datasets keep increasing in size, computational complexity and communication cost of decentralized optimization start presenting major challenges. To reduce the complexity of computing gradients, decentralized stochastic methods that allow each agent to perform gradient estimation by processing small subset of local data are preferred \cite{ram2010distributed}. However, such techniques exhibit low convergence rates and suffer from high variance of the local stochastic gradients. 
Remedies for these impediments in decentralized optimization over networks and in stochastic optimization in centralized settings include
gradient tracking \cite{qu2017harnessing, nedic2017achieving} and variance reduction 
\cite{johnson2013accelerating, schmidt2017minimizing}, respectively; however, no such remedies have been developed for decentralized optimization over time-varying directed networks.
On another note, communication constraints often exacerbate large-scale decentralized optimization problems where the size of the network or the dimension of local model parameters may be on the order of millions. This motivates design of communication-efficient algorithms that compress messages exchanged between network nodes yet preserve fast convergence. In this paper, we study the general setting where a network is time-varying and directed, and present, to the best of our knowledge, the first variance-reduced communication-sparsified algorithm for decentralized convex optimization over such networks. Moreover, we theoretically establish that when the local objective functions are smooth and strongly-convex, the proposed algorithm achieves accelerated linear convergence rate. 
Note that while our focus is on communication-constrained settings, the proposed scheme readily applies to decentralized optimization problems where time-varying directed graphs operate under no communication constraints; in fact, to our knowledge, this is the first variance-reduced stochastic algorithm for such problems. 

\subsection{Related work}


The first work on decentralized optimization over networks dates back to the 1980s \cite{tsitsiklis1984problems}. A number of studies that followed in subsequent years was focused on the decentralized average consensus problem, where the network nodes work collaboratively to compute the average value of local vectors. Convergence conditions for achieving consensus over directed and undirected time-varying graphs were established in \cite{jadbabaie2003coordination,ren2005consensus,ren2007information,cai2012average,cai2014average}. The first communication-efficient algorithm that achieves linear convergence over time-invariant (static) undirected graphs was proposed in \cite{koloskova2019decentralized}.

The consensus problem can be viewed as a stepping stone towards more general decentralized optimization problems, where the nodes in a network aim to collaboratively minimize the sum of local objective functions. A number of solutions to this problem has been proposed for the setting where the network is undirected, including the well-known distributed (sub)gradient descent algorithm (DGD) \cite{nedic2009distributed, johansson2010randomized}, distributed alternating direction method of multipliers (D-ADMM) \cite{wei2012distributed}, and decentralized dual averaging methods \cite{duchi2011dual, nedic2015decentralized, he2018cola}. Recently, \cite{stich2018sparsified,koloskova2019decentralized} proposed a novel communication-efficient algorithm for decentralized convex optimization problems; the provably convergent algorithm relies on a message-passing scheme with memory and biased compression.


A key technical property required to ensure convergence of decentralized convex optimization algorithms over undirected networks is that the so-called mixing matrix characterizing the network connectivity is doubly-stochastic. However, in directed networks characterized by communication link asymmetry, doubly-stochastic mixing matrices are atypical. This motivated algorithms that rely on auxiliary variables to cancel out the imbalance in asymmetric directed networks in order to achieve convergence. For instance, the subgradient-push algorithm \cite{kempe2003gossip,nedic2014distributed} works with column-stochastic mixing matrices and introduces local normalization scalars to ensure converge. The directed distributed gradient descent (D-DGD) algorithm
\cite{xi2017distributed}, on the other hand, keeps track of the variations of local models by utilizing auxiliary variables of the same dimension as the local model parameters. For convex objective functions, both algorithms achieve $\O(\frac{\mathrm{ln}T}{\sqrt{T}})$ convergence rate. When the objectives are strongly-convex with Lipshitz gradients, and assuming availability of only the stochastic gradient terms, the stochastic gradient-push algorithm proposed in \cite{nedic2016stochastic} achieves $\O(\frac{\mathrm{ln}T}{T})$ convergence rate. A common feature of these algorithms is their reliance upon diminishing stepsizes to achieve convergence to the optimal solution; in comparison, using a fixed stepsize can accelerate the gradient search but cannot guarantee the exact convergence, only the convergence to a neighborhood of the optimal solution. The implied exactness-speed dilemma can be overcome using schemes that deploy gradient tracking (see, e.g., \cite{qu2017harnessing, nedic2017achieving, saadatniaki2018optimization}). These schemes utilize fixed step sizes to achieve linear convergence rate when the objective functions are both smooth and strongly-convex. Among them, the Push-DIGing algorithm \cite{nedic2017achieving} follows the same basic ideas of the subgradient-push algorithm, while TV-AB \cite{saadatniaki2018optimization} relies on column- and row-stochastic matrices to update model parameters and gradient terms, respectively. 


The aforementioned linearly convergent methods rely on full gradient, i.e., each node is assumed to use all of its data to compute the local gradient. However, if the number of data points stored at each node is large, full gradient computation becomes infeasible. To this end, stochastic gradient descent was adapted to decentralized settings, but the resulting computational savings come at the cost of sublinear convergence rate \cite{xin2019variance, nedic2016stochastic}. To accelerate traditional stochastic gradient methods in centralized settings, variance-reduction algorithms such as SAG \cite{schmidt2017minimizing} and SVRG \cite{johnson2013accelerating} have been proposed; these schemes enable linear convergence when the objective functions are smooth and strongly-convex. In decentralized settings, GT-SVRG \cite{xin2019variance} and Network-SVRG \cite{li2020communication} leverage variance-reduction techniques to achieve linear convergence rate. However, these algorithms are restricted to static and undirected networks, and a narrow class of directed networks where the mixing matrices can be rendered doubly stochastic.\footnote{To have doubly stochastic mixing matrices, directed graphs require weight balance, i.e., at each node of a graph, the sum of the weights from in-coming edges should be equal to that of the weights from out-coming edges \cite{gharesifard2010does}.} The existing algorithms for decentralized optimization over directed networks, such as Push-SAGA \cite{qureshi2021push}, are restricted to static networks.

In recent years, decentralized learning tasks have experienced rapid growth in the amount of data and the dimension of the optimization problems, which may lead to practically infeasible demands for communication between network nodes. To this end, various communication compression schemes have been proposed; among them, the most frequently encountered are quantization and sparsification. Quantization schemes limit the number of bits encoding the messages, while the sparsification schemes select a subset of features and represent messages in lower dimensional space. For instance, \cite{tang2018communication, wen2017terngrad, zhang2017zipml,stich2018sparsified, das2020improved, reisizadeh2019robust} propose algorithms for distributed training of (non)convex machine learning models in static master-worker settings (i.e., star graph topology) using quantized/compressed information, while \cite{shen2018towards,koloskova2019decentralized,koloskova2019decentralizeda,hashemi2020delicoco} develop communication-efficient algorithms for decentralized optimization over static and undirected networks. However, directed networks in general, and time-varying ones in particular, have received considerably less attention. Decentralized optimization over such networks faces technical challenges of developing an algorithmic framework conducive to theoretical analysis and establishing convergence guarantees, which is further exacerbated when the communication is compressed. Early steps in this direction were made in \cite{taheriquantized} by building upon the subgradient-push algorithm to develop a quantized communication framework for decentralized optimization over a static network. 

Our proposed algorithm utilizes gradient tracking and variance reduction to achieve fast convergence, and relies on stochastic gradients to solve the decentralized convex optimization task at feasible computational cost. Preliminary results of this work, focused on a significantly slower full gradient framework ($\O(\frac{1}{\epsilon^2})$ vs. $\O(\ln \frac{1}{\epsilon})$), were presented in \cite{chen2020communication}.
In Table \ref{tb:comp}, we briefly summarize and contrast several algorithms for decentralized optimization over directed graphs.

\subsection{Notation}
We use lowercase bold letters to represent vectors and uppercase letters to represent matrices. $ [A]_{ij}$ denotes the $(i, j)$ entry of matrix $A$, while $\| \cdot \|$ denotes the standard Euclidean norm. 
The spectral radius of a matrix $A$ is denoted by $\rho(A) $.
The weighted infinity norm of $\x$ given a positive vector $\mathbf{w}$ is $\|\x \|^{\mathbf{w}}_{\infty} = \max_i |x_i|/w_i$ and the induced matrix norm is $ \| |\cdot | \|_{\infty}^{\mathbf{w}}$. Finally, $I$ denotes the identity matrix whose dimension is inferred from the context.

\begin{table*}[t]
\caption{The settings and convergence rates of algorithms for decentralized optimization over directed graphs.
}
\centering
\ra{1}
\scriptsize 
\begin{tabular*}{.99\linewidth}{@{}cccccc@{}}\toprule
Algorithm &Convergence& Digraph & Gradient & Convex Objective Setting&Compress\\ \midrule
Subgradient-push \cite{nedic2016stochastic} & $\O(\frac{1}{\epsilon})$& 
Time-varying & Stochastic & Strong convexity
&{\color{red}\xmark}\\\midrule
Push-DIGing \cite{nedic2017achieving} & $\O(\ln \frac{1}{\epsilon})$& Time-varying &Full & Strong convexity and smoothness &{\color{red}\xmark}\\\midrule
TV-AB \cite{saadatniaki2018optimization} & $\O(\ln \frac{1}{\epsilon})$& Time-varying & Full & Strong convexity and smoothness&{\color{red}\xmark}\\\midrule
Quantized Push-sum \cite{taheriquantized} & $\O(\frac{1}{\epsilon^2})$& Static & Full &  -&{\color{black}\cmark}\\\midrule
{\bf This work} & $\O(\ln \frac{1}{\epsilon})$& Time-varying & Stochastic & Strong convexity and smoothness &{\color{black}\cmark}\\
\bottomrule
\end{tabular*}
\label{tb:comp}
\end{table*}

\section{Preliminaries}\label{sec:pre}
\subsection{Problem Formulation}\label{sec21}

For convenience, we remind the reader of the problem formulation (\ref{eq:problem}): In a network of $n$ agents, where each node maintains a local model consisting of $d$ parameters, the agents  collaboratively solve the decentralized convex optimization
\begin{equation} \label{eq:prob}
\min_{\mathrm{\mathbf{x} \in \mathbb{R}^d}} \left[f(\mathbf{x}):=\frac{1}{n}\sum_{i=1}^n f_i(\mathbf{x})\right],
\end{equation}
where $f_i(\mathbf{x}) = \frac{1}{m_i}\sum_{j=1}^{m_i}f_{i, j}(\mathbf{x}): \mathbb{R}^d \to \mathbb{R}$ for $i \in [n]:=\left\{1, ..., n \right\}$ denotes the local objective function at node $i$. Each component of $f_i$ is assumed to be smooth and not accessible to nodes other than the $i^{th}$ one, and the global objective $f$ is assumed to be strongly-convex. We further assume existence of a unique optimal solution $\x^* \in \mathbb{R}^d $ and that each node can communicate to its neighbors; the nodes identify $\x^*$ by exchanging messages over a time-varying directed network. The network's connectivity properties are elaborated upon in Section~3. 


\subsection{Communication-Efficient Methods}\label{sec22}

In practice, bandwidth limitations may restrict the amount of data that the network nodes can communicate to each other; this is typical of high-dimensional scenarios where the dimension $d$ of local parameters $\x_i$ is exceedingly large. To handle communication constraints, network nodes may employ {\it sparsification} to reduce the size of the messages. Typically, there are two approaches to sparsification: (i) each node selects and communicates $d_q$ out of $d$ components of a $d$-dimensional message; or (ii) each component of a $d$-dimensional message is selected to be communicated independently with probability $d_q/d$. Note that the former imposes a hard constraint on the number of communicated entries while the latter results in $d_q$ communicated entries on expectation; both select a specific entry with probability $d_q/d$. Throughout this paper, we focus on and study the first approach.

Let $Q: \mathbb{R}^d \rightarrow \mathbb{R}^d$ denote the sparsification operator; we allow for biased $Q$ with variance proportional to the argument norm, i.e., $\mathbb{E}[Q(\x)] \neq \x$ and $\mathbb{E}[\|Q(\x)-\x\|^2] \propto \|\x\|^2$. This stands in contrast to typical compression operators which aim to achieve no bias and have bounded variance (see, e.g., \cite{tang2018communication}). More recent works \cite{stich2018sparsified,koloskova2019decentralized,koloskova2019decentralizeda,taheriquantized} do consider biased compression operators but only for time-invariant communication networks -- a setting that is more restrictive than the one considered in this paper.


\section{Algorithm development}\label{sec:alg}
In this section, we first introduce a novel average consensus algorithm, an intermediate step towards the main (optimization) framework; then we present the optimization algorithm consisting of the consensus and the gradient components.

\subsection{Foundations: Decentralized average consensus}

We start by specifying a procedure for decentralized average consensus, an intermediate step towards decentralized optimization and, ultimately, an integral part thereof. The decentralized average consensus problem is formulated as the computation $\bar{\x} = \frac{1}{n}\sum_{i=1}^n \x_i$, where $\x_i \in \mathcal{R}^d$ is the parameter vector at node $i$. Following the idea of \cite{cai2012average}, for each node of the network we define the so-called surplus vector, i.e., an auxiliary variable $\mathbf{y}_i\in \mathbb{R}^d$ which tracks local state vector variations over consecutive time steps; as shown later in this section, one can use the surplus vector to help provide guarantees of convergence to the optimal solution of the decentralized problem. The surplus vector is exchanged along with the state vector, i.e., at time $t$, node $i$ sends both $\mathbf{y}_i^t$ and the state vector $\mathbf{x}_i^t$ to its out-neighbors. For the sake of having compact notation, let us introduce $\mathbf{z}_i^t \in \mathbb{R}^d$,
\begin{equation}\label{eq:zdef}
\mathbf{z}_i^t = \begin{cases}
\mathbf{x}_i^t, & i \in \left\{1, ..., n \right\} \\
\mathbf{y}_{i-n}^t, & i \in \left\{n+1, ..., 2n \right\},
\end{cases}
\end{equation}
to represent messages node $i$ communicates to its neighbors in the network at time $t$. 

We assume that the time-varying graph is $\mathcal{B}$-jointly connected, i.e., that there exists a window size $\mathcal{B}\geq1$ such that the aggregate graph $\bigcup_{l=t}^{t+\mathcal{B}-1}\G_{l}$ is strongly connected for all $t = k\mathcal{B}$, $k \in \mathcal{N}$. Note that if $\mathcal{B}=1$, each instance of the graph is strongly connected. This is a more general assumption than the often used $\mathcal{B}$-bounded strong-connectivity (see, e.g. \cite{saadatniaki2018optimization}) which requires strong connectivity of the union graph $\bigcup_{l=t}^{t+\mathcal{B}-1}\G_{l}$ for all $t \geq 0$.

For $\mathcal{B}$-jointly connected graphs, the product of mixing matrices of graph instances over $\mathcal{B}$ consecutive time steps has a non-zero spectral gap.
To formalize this statement, let us construct two weight matrices that reflect the network topology; in particular, let $W_{in}^t$ (row-stochastic) and $W_{out}^t$ (column-stochastic) denote the in-neighbor and out-neighbor weight matrices at time $t$, respectively. It holds that $[W_{in}^t]_{ij}>0 $ if and only if $ j \in \mathcal{N}_{{in},i}^t$ and $[W_{out}^t]_{ij} >0 $ if and only if $i \in \mathcal{N}_{out,j}^t$,
where $\mathcal{N}_{{in},i}^t$ denotes the set of nodes that may send information to node $i$ (including $i$) whereas
$\mathcal{N}_{out,j}^t$ denotes the set of nodes that may receive information from node $j$  (including $j$) at time 
$t$. We assume $W_{in}^t$ and $W_{out}^t$ are given and that both $\mathcal{N}_{{in},i}^t$ and 
$\mathcal{N}_{out,i}^t$ are known to node $i$. A common policy for designing $W^t_{in}$ and $W^t_{out}$ is to assign
\begin{equation}\label{eq:Ws}
[W^t_{in}]_{ij} = 1/|\mathcal{N}^t_{in, i}|, \qquad [W^t_{out}]_{ij}= 1/|\mathcal{N}^t_{out, j}|.
\end{equation}

Recall that we are interested in sparsifying messages exchanged between nodes of a network; clearly, the sparsification should impact the structure of a mixing matrix. Indeed, if one attempts sparsifying messages used by existing methods, e.g. \cite{kempe2003gossip,cai2012average,cai2014average,nedic2014distributed}, without any modifications of the mixing matrices therein, non-vanishing error terms induced by the compression operator will prevent those methods from converging. We note that the impact of sparsification on the components of a message vector is similar to the impact of link failures, and may therefore be captured by the structure of the weight matrices. To elaborate on this, observe that the vector-valued problem at time $t$ can essentially be decomposed to $d$ individual scalar-valued tasks with weight matrices $\{W_{in,m}^t\}_{m=1}^d$ and $\{W_{out,m}^t\}_{m=1}^d$. For the sparsified components of the message vector, i.e., those that are set to zero and not communicated, the corresponding entries in the weight matrices can be replaced by zeros; on the other hand, the entries in the weight matrices corresponding to the communicated components of the message vector remain unchanged, leading to the violation of the stochasticity of the weight matrices. To address this, we {\it re-normalize} the weight matrices $\{W_{in,m}^t\}_{m=1}^d$ and $\{W_{out,m}^t\}_{m=1}^d$, thus ensuring their row and column stochasticity. Note that the re-normalization of the $i\ts{th}$ row of $\{W_{in,m}^t\}_{m=1}^d$ ($i\ts{th}$ column of $\{W_{out,m}^t\}_{m=1}^d$) is performed by the $i^{th}$ network node. 

To specify the normalization rule, we first need to define the sparsification operation. Sparsification of $\mathbf{x}_i^t$ (and, consequently, $\mathbf{y}_i^t$) is done via the compression operator $Q(\cdot)$ applied to $\mathbf{z}_i^t$; we denote the result by $Q(\mathbf{z}_i^t)$. Let $[Q({\z}_{i}^t)]_m$ denote the $m\ts{th}$ component of $Q(\mathbf{z}_i^t)$. 
Let $\{A_{m}^t\}_{m=1}^d$ and $\{B_{m}^t\}_{m=1}^d$ be the weight matrices obtained after normalizing $\{W_{in,m}^t\}_{m=1}^d$ and $\{W_{out,m}^t\}_{m=1}^d$, respectively. To formalize the normalization procedure, we introduce the weight matrix
\begin{equation}\label{eq:normA}
[A^t_m]_{ij}=\begin{cases}
\frac{[W^t_{in, m}]_{ij}}{\sum_{j\in\mathcal{S}_m^t(i,j)} [W^t_{in, m}]_{ij}} & \text{if } j\in\mathcal{S}_m^t(i,j)\\
0 & \mathrm{otherwise},
\end{cases}
\end{equation}
where $\mathcal{S}_m^t(i,j) := \{j|j\in\mathcal{N}^t_{in,i},[Q({\z}_{j}^t)]_m \neq 0\}\cup\{i\}$. Likewise, the weight matrix $B^t_m$ is defined as
\begin{equation}\label{eq:normB}
[B^t_m]_{ij}=\begin{cases}
\frac{[W^t_{out, m}]_{ij}}{\sum_{i\in\mathcal{T}_m^t(i,j)} [W^t_{out, m}]_{ij}} & \text{if } i\in\mathcal{T}_m^t(i,j)\\
0 & \mathrm{otherwise},
\end{cases}
\end{equation}
where $\mathcal{T}_m^t(i,j) := \{i|i\in\mathcal{N}^t_{out,j},[Q({\z}_{i}^t)]_m \neq 0\}\cup\{j\}$.

We can now define the {\it mixing matrix} of a directed network with sparsified messages. 

\begin{definition}
The $m\ts{th}$ mixing matrix at time $t$ of a time-varying directed network with sparsified messages, $\Bar{M}_m^t \in \mathbb{R}^{2n \times 2n}$, is a matrix whose columns sum up to $1$ and whose eigenvalues satisfy $1=|\lambda_1(\Bar{M}_m^t)| = |\lambda_2(\Bar{M}_m^t)| \geq |\lambda_3(\Bar{M}_m^t)| \geq \cdots |\lambda_{2n}(\Bar{M}_m^t)|$, constructed from the current network topology as
\begin{equation}\label{matrixdefbar}
\Bar{M}_m^t=\left[\begin{matrix}
   A_m^t & \mathbf{0} \\
   I-A_m^t & B_m^t \\
  \end{matrix}
\right],
\end{equation}
where $A_m^t$ and $B_m^t$ denote the $m\ts{th}$ normalized in-neighbor and out-neighbor weight matrices at time $t$, respectively. 
\end{definition}

Given $\mathbf{z}_i^t$ and $\bar{M}_m^t$ in \eqref{eq:zdef} and \eqref{matrixdefbar}, respectively, we can formulate a compact recursive update rule for $\mathbf{z}_i^t$ as
\begin{equation}\label{eq:sparsify_cons_update}
\begin{aligned}
z_{im}^{t+1} &=\sum_{j=1}^{2n}[\bar{M}^t_m]_{ij} [Q(\z_{j}^t)]_m + \mathbbm{1}_{\left\{t\ \text{mod}\ \mathcal{B} = \mathcal{B}-1\right\}} \gamma [F]_{ij} z_{jm}^{\mathcal{B} \lfloor t/\mathcal{B} \rfloor}, \\
\end{aligned}
\end{equation}
where $F =\left[ \begin{matrix} \mathbf{0} & I \\
\mathbf{0} & -I 
\end{matrix} \right]$ and $m$ denotes the coordinate index.

As seen in \eqref{eq:sparsify_cons_update}, vectors $\z_i^t$ (which contain $\x_i^t$, the quantities to be averaged) are updated in a straightforward manner via sparsification and multiplication with the mixing matrix at all times $t$ except those that satisfy
\begin{equation}\label{eq:mod}
    t \mod \mathcal{B} = \mathcal{B}-1.
\end{equation}
In particular, when (\ref{eq:mod}) holds, vectors $\z_i^{\mathcal{B} \lfloor t/\mathcal{B} \rfloor}$, stored at time $\mathcal{B} \lfloor t/\mathcal{B} \rfloor $, are also used to update $\z_i^t$. The usage of the stored vectors is motivated by the observation that $\bar{M}_m^t$ may have zero spectral gap, which is undesirable since for such mixing matrices the convergence of the consensus algorithms will not be guaranteed. However, for a judiciously chosen perturbation parameter $\epsilon$, which determines
to what extent $\sum_{j=1}^{2n}[F]_{ij} z_{jm}^{\mathcal{B} \lfloor t/\mathcal{B} \rfloor}$ affects the update, we can ensure a nonzero spectral gap of the product of $\mathcal{B}$ consecutive mixing matrices starting from $t = k\mathcal{B}$. The described communication-sparsified average consensus procedure over directed graphs, referred to for convenience as Di-CS-AC, is formalized as Algorithm \ref{alg:A}.

\begin{algorithm}[t]
\caption{Directed Communication-Sparsified Average Consensus (Di-CS-AC)}
\label{alg:A}
\begin{algorithmic}[1]
\STATE {\bfseries Input:} 
   $T$, 
   $\mathbf{x}^0$, 
   $\mathbf{y}^0=\mathbf{0}$, 
   $\gamma$
   
\STATE {set $\mathbf{z}^0=[\mathbf{x}^0; \mathbf{y}^0]$,  $\Tilde{w}^0 = \z^0 $ and } 
\FOR{each $s \in [0, 1, ..., S]$}

\STATE generate non-negative matrices $\{W_{in,m}^t\}_{m=1}^d$ and $\{W_{out,m}^t\}_{m=1}^d$
\FOR{each $m \in [1, ..., d]$}
\STATE construct a row-stochastic $A^t_m$  and a column-stochastic $B^t_m$ according to \eqref{eq:normA} and \eqref{eq:normB}
\STATE construct $\bar{M}^t_m$ according to \eqref{matrixdefbar}
\FOR{each $i \in [1, ..., 2n ]$}
\STATE update $z_{im}^{t+1}$ according to \eqref{eq:sparsify_cons_update}
\ENDFOR
\ENDFOR

\ENDFOR

\end{algorithmic}
\end{algorithm}

\subsection{Decentralized gradient component }

Going beyond the simple consensus problem and towards solving optimization \eqref{eq:prob}, we re-define the recursive update rule for $\mathbf{z}_i^t$ as
\begin{equation}\label{eq:sparsify_vr_update}
\begin{aligned}
z_{im}^{t+1} &=\sum_{j=1}^{2n}[\bar{M}^t_m]_{ij} [Q(\z_{j}^t)]_m + \mathbbm{1}_{\left\{t\ \text{mod}\ \mathcal{B} = \mathcal{B}-1\right\}} \gamma [F]_{ij} z_{jm}^{\mathcal{B} \lfloor t/\mathcal{B} \rfloor}  - \mathbbm{1}_{\left\{t\ \text{mod}\ \mathcal{B} = \mathcal{B}-1\right\}} \alpha g_{im}^{\mathcal{B} \lfloor t/\mathcal{B} \rfloor},
\end{aligned}
\end{equation}
where $F$ and $m$ denote the same objects as in (\ref{eq:sparsify_cons_update}), and $g_{im}$ combines global gradient tracking with local stochastic reduction to achieve accelerated convergence (elaborated upon shortly). Note that \eqref{eq:sparsify_vr_update} implies the following element-wise update rules for state and surplus vectors, respectively:
\begin{equation}\label{update_consensus_x}
\begin{aligned}
x_{im}^{t+1} & = \sum_{j=1}^{n}[A_m^t]_{ij}[Q(\x_j^t)]_m + \mathbbm{1}_{\left\{t\ \text{mod}\ \mathcal{B} = \mathcal{B}-1\right\}} \gamma  y_{im}^{\mathcal{B} \lfloor t/\mathcal{B} \rfloor}  - \mathbbm{1}_{\left\{t\ \text{mod}\ \mathcal{B} = \mathcal{B}-1\right\}} \alpha g_{im}^{\mathcal{B} \lfloor t/\mathcal{B} \rfloor},
\end{aligned}
\end{equation}
\begin{equation}\label{update_consensus_y}
\begin{aligned}
    y_{im}^{t+1} & = \sum_{j=1}^{n}[B_m^t]_{ij}[Q(\y_j^t)]_m   - (x_{im}^{t+1} - x_{im}^t). 
\end{aligned}
\end{equation}

Paralleling the basic consensus task discussed in the previous section, vectors $\z_i^t$ (containing state vectors to be averaged) are updated via sparsification and multiplication with the mixing matrix at all times $t$ except those that satisfy \eqref{eq:mod}. When (\ref{eq:mod}) does hold, vectors $\z_i^{\mathcal{B}\lfloor t/\mathcal{B} \rfloor}$, stored at times $\mathcal{B}\lfloor t/\mathcal{B} \rfloor$, are also used to update $\z_i^t$; the motivation and reasoning for such special treatment are as same as in the consensus algorithm.\footnote{Note that $F$ has all-zero matrices for its $(1, 1)$ and $(2, 1)$ blocks and thus we only need to store $\z_i^{\mathcal{B}\lfloor t/\mathcal{B} \rfloor}$ (equivalently, $\y_{i-n}^{\mathcal{B}\lfloor t/\mathcal{B} \rfloor}$), where $n+1 \leq i \leq 2n$.}

In the proposed algorithm, updates of the gradient term $\g_i^t$ combine global gradient tracking with local stochastic variance reduction. 
In particular, the updates of $\g_i^t$ mix gradient messages while keeping track of the changes in gradient estimates $\v_i^t$; this guides $\g_i^t$ towards the gradient of the global objective, ultimately ensuring convergence to the optimal solution $\mathbf{x}^*$ (i.e., global gradient tracking helps avoid the pitfall of non-vanishing local gradients which would otherwise lead the search only to a neighborhood of $\mathbf{x}^*$). The $m\ts{th}$ entry of $\g_i^t$, $g_{im}^t$, is updated as
\begin{equation}
\begin{aligned}
    g_{im}^{\mathcal{B} (\lfloor t/\mathcal{B} \rfloor)} = 
\begin{cases}
\sum_{j=1}^{n}[B_m(k\mathcal{B}-1:(k-1)\mathcal{B})]_{ij} g_{jm}^{(k-1)\mathcal{B}} + v_{im}^{\mathcal{B} (\lfloor t/\mathcal{B} \rfloor)}-v_{im}^{\mathcal{B} (\lfloor t/\mathcal{B} \rfloor -1)}  & i \leq n \\
0 & \mathrm{otherwise},
\end{cases}
\end{aligned}
\label{gradg}
\end{equation}
where $k = \lfloor t/\mathcal{B} \rfloor$. 
The gradient estimate $\v_i^t$ in (\ref{gradg}) is updated via the stochastic variance-reduction method \cite{johnson2013accelerating}. Specifically,
\begin{equation}\label{eq:update_v}
    \mathbf{v}_i^{\mathcal{B} (\lfloor t/\mathcal{B} \rfloor+1)}=\nabla f_{i, l_i}(\z_i^{\mathcal{B} \lfloor t/\mathcal{B} \rfloor})-\nabla f_{i, l_i}(\Tilde{w}_{i})+\Tilde{\mu}_i, \quad \forall i \in [n].
\end{equation}
One can interpret this update as being executed in a double loop fashion: when a local full gradient at node $i$, $\Tilde{\mu}_i$, is computed (in what can be considered an outer loop), it is retained in the subsequent $T$ iterations (the inner loop). In each iteration of the inner loop, if the time step satisfies \eqref{eq:mod}, node $i$ uniformly at random selects a local sample, $l_i$, for the calculation of two stochastic gradient estimates -- an estimate of the current state, $\nabla f_{i, l_i}(\z_i^{\mathcal{B} \lfloor t/\mathcal{B} \rfloor})$, and an estimate of the state from the last outer loop, $ \nabla f_{i, l_i}(\Tilde{w}_{i})$ -- the terms needed to perform update of $\mathbf{v}_i^t$. By computing a full gradient periodically in the outer loop and estimating gradient stochastically in the inner loop, the described procedure trades computational cost for convergence speed, ultimately achieving linear convergence at fewer gradient computations per sample than the full gradient techniques.

The described procedure is formalized as Algorithm \ref{alg:B}.

\begin{remark}
We highlight a few important observations regarding Algorithm \ref{alg:B}.
\begin{enumerate}[(a)]
\item When there are no communication constraint and each agent in the network can send full information to out-neighboring agents, Algorithm \ref{alg:B} reduces to a novel stochastic variance-reduced scheme for decentralized convex optimization over such networks. 
\item For $\mathcal{B}=1$, the problem reduces to decentralized optimization over networks that are strongly connected at all time steps, a typical connectivity assumption for many decentralized optimization algorithms \cite{xi2017distributed, koloskova2019decentralized}. 
\item  Algorithm \ref{alg:B} requires each node in the network to store local vectors of size $4d$, including the current state vector, current and past surplus vector, and local gradient vector. While the current state vector and current surplus vector may be communicated to the neighboring nodes, past surplus vectors are only used locally to add local perturbations at the time steps satisfying \eqref{eq:mod}. 
\item The columns of $\bar{M}_m^t$ sum up to one. However, $\bar{M}_m^t$ is not column-stochastic as it has negative entries, which stands in contrast to the stochasticity property of the mixing matrices appearing in the average consensus algorithms \cite{xiao2004fast,koloskova2019decentralized}. 
\end{enumerate}
\end{remark}



\begin{algorithm}[t]
\caption{Directed Communication-Sparsified Stochastic Variance-Reduced Gradient Descent (Di-CS-SVRG)}
\label{alg:B}
\begin{algorithmic}[1]
\STATE {\bfseries Input:} 
   $T$, 
   $\mathbf{x}^0$, 
   $\mathbf{y}^0=\mathbf{0}$, 
   $\alpha$, 
   $\gamma$
   
\STATE {set $\mathbf{z}^0=[\mathbf{x}^0; \mathbf{y}^0]$,  $\Tilde{w}^0 = \z^0 $ and $\g_i^0 = \v_i^0 = \nabla \f_i(\x_i^0) \quad \forall i \in [n] $} 
\FOR{each $s \in [0, 1, ..., S]$}
\STATE $\Tilde{w}=\Tilde{w}^{s}$
\STATE $\Tilde{\mu}_i=\nabla f_i(\Tilde{w}) =  \frac{1}{m_i}\sum_{j=1}^{m_i} \nabla f_{i, j}(\Tilde{w})$
\FOR{each $t \in [sT + 1,..., (s+1)T - 1]$}
\STATE generate non-negative matrices $\{W_{in,m}^t\}_{m=1}^d$ and $\{W_{out,m}^t\}_{m=1}^d$
\FOR{each $m \in [1, ..., d]$}
\STATE construct a row-stochastic $A^t_m$  and a column-stochastic $B^t_m$ according to \eqref{eq:normA} and \eqref{eq:normB}
\STATE construct $\bar{M}^t_m$ according to \eqref{matrixdefbar}
\FOR{each $i \in [1, ..., 2n ]$}
\STATE update $z_{im}^{t+1}$ according to \eqref{eq:sparsify_vr_update}
\ENDFOR
\IF{ $t\mod \mathcal{B} = \mathcal{B}-1 $}
\FOR{each $i \in [1, ..., n ]$}
\STATE \quad select $l_i$ uniformly randomly from $[m_i]$:

\STATE \qquad update $\mathbf{v}_i^{\mathcal{B} (\lfloor t/\mathcal{B} \rfloor+1)} $ according to \eqref{eq:update_v}
\STATE \qquad update $g_{im}^{\mathcal{B} (\lfloor t/\mathcal{B} \rfloor+1)} 
$ according to \eqref{gradg} 
\ENDFOR
\ENDIF
\ENDFOR
\ENDFOR
\STATE $\Tilde{w}^{s+1}=\z^{(s+1)T}$
\ENDFOR

\end{algorithmic}
\end{algorithm}

\section{Convergence Analysis}\label{sec:thm}
For convenience, let us denote the product of a sequence of mixing matrices from time step $s$ to $T$ as 
\begin{equation}
 \bar{M}_m(T:s) = \bar{M}_m^{T} \bar{M}_m^{T-1} \cdots \bar{M}_m^s.
\end{equation}
To further simplify notation, we also introduce
\begin{equation}\label{mproduct1}
M_m((k+1)\mathcal{B}-1:k\mathcal{B}) = \bar{M}_m((k+1)\mathcal{B}-1:k\mathcal{B}) + \gamma F, \end{equation}
and
\begin{equation}
\begin{aligned}
M_m(t:k_1\mathcal{B})
= \bar{M}_m(t:k_2\mathcal{B})M_m(k_2\mathcal{B}-1:(k_2-1)\mathcal{B}) \cdots 
M_m((k_1+1)\mathcal{B}-1:k_1\mathcal{B}),
\end{aligned}
\end{equation}
where $k_2\mathcal{B} \leq t \leq (k_2+1)\mathcal{B}-1$ and $k_1, k_2 \in \mathcal{N}, k_1 \leq k_2$. Note that
$M_m((k+1)\mathcal{B}-1:k\mathcal{B})$ is formed by adding a perturbation matrix $\gamma F $ to the product $\bar{M}_m((k+1)\mathcal{B}-1:k\mathcal{B})$.
Finally, we also introduce shorthand notation for the product of the weight matrices $B_m$ from time $s$ to $T$,
\begin{equation}
B_m(T:s) = B_m^{T} B_m^{T-1} \cdots B_m^s.
\end{equation}

Our analysis relies on several standard assumptions about the graph and network connectivity matrices as well as the characteristics of the local and global objective functions. These are given next.
\begin{assumption}\label{new_assp} 
Suppose the following conditions hold:
\begin{enumerate}[(a)]
    \item The product of consecutive mixing matrices $M_m((k+1)\mathcal{B}-1:k\mathcal{B})$ in (\ref{mproduct1}) has a non-zero spectral gap for all $k \geq 0$, $1 \leq m \leq d$, and all $0 < \gamma < \gamma_0$ for some $0< \gamma_0 <1$. \label{assumption5}
    \item The collection of all possible mixing matrices $\{\bar{M}_m^t\}$ is a finite set.
    \label{assumption6}
\item Each component of the local objective function $f_{i, j}$ is $L$-smooth
and the global objective $f$ is $\mu$-strongly-convex\footnote{This implies that
$\forall \x_1, \x_2 \in \R^d$ there exists $L >0$ such that
    $\|\nabla f_{i, j}(\x_1) - \nabla f_{i, j}(\x_2) \| \leq L\|\x_1 - \x_2 \|$.
Furthermore, $\forall \x_1, \x_2 \in \R^d$ there exists $\mu >0$ such that
    $f(\x_2) \geq f(\x_1) + \langle \nabla f(\x_1), \x_2 - \x_1 \rangle + \frac{\mu}{2} \|\x_2 - \x_1 \|^2$.
}.
\end{enumerate}
\end{assumption}
\begin{remark}
Assumption \ref{new_assp}(\ref{assumption5}) is readily satisfied for a variety of graph structures such as the $\mathcal{B}$-strongly connected directed graph introduced in \cite{nedic2017achieving}, i.e., the setting where the union of graphs over $B$ consecutive instances starting from $k\mathcal{B}$ forms a strongly connected graph for any non-negative integer $k$.\footnote{There are two versions of the definition of $\mathcal{B}$-strongly connected directed graphs, the difference being the window starting time. As noted in Section~II, we consider the definition where the window may start at any time $t = k\mathcal{B}$; this differs from the (more demanding in regards to connectivity) definition in \cite{saadatniaki2018optimization} where the starting time is an arbitrary non-negative integer.} Furthermore, one can readily verify that Assumption \ref{new_assp}(\ref{assumption6}) holds for the weight matrices defined in \eqref{eq:Ws}.
\end{remark}


Before stating the main theorem, we provide the following lemma which, under Assumption \ref{new_assp}, establishes the consensus contraction of the product of mixing matrices and the product of normalized weight matrices.


\begin{lemma}\label{lemma1}
Suppose Assumptions  \ref{new_assp}(\ref{assumption5}) and \ref{new_assp}(\ref{assumption6}) hold. Let $\sigma = \max(|\lambda_{M, 2}|, |\lambda_{B, 2}|)$ denote the larger of the second largest eigenvalues of $M_m((k+1)\mathcal{B}-1:k\mathcal{B} )$ and $B_m((k+1)\mathcal{B}-1:k\mathcal{B})$. Then, 
\begin{align}
\begin{split}
    \|M_m((k+1)\mathcal{B}-1:k\mathcal{B} ) \z - \Bar{\z} \| \leq \sigma \|\z - \Bar{\z} \|, \ \forall \z \in \R^{2n} \\
    \mbox{and}
    \\
    \|B_m((k+1)\mathcal{B}-1:k\mathcal{B} ) \y - \Bar{\y} \| \leq \sigma \|\y - \Bar{\y} \|, \ \forall \y \in \R^{n}, \\
\end{split}
\end{align}
where $\Bar{\z} = [\frac{1}{n}\sum_{i=1}^{2n}z_i, \cdots, \frac{1}{n}\sum_{i=1}^{2n}z_i]^T $ and $\Bar{\y} = [\frac{1}{n}\sum_{i=1}^{n}y_i, \cdots, \frac{1}{n}\sum_{i=1}^{n}y_i]^T $. 
\end{lemma}
\begin{proof}
The proof of the lemma is in the supplementary document.
\end{proof}
Our main result, stated in Theorem \ref{thm:main} below, establishes that Algorithm \ref{alg:B} provides linear convergence of local parameters to their average values, which itself converges linearly to the optimal solution of \eqref{eq:problem}.
\begin{theorem}\label{thm:main}
Suppose Assumption \ref{new_assp} holds. 
Denote the condition number of $f$ by $\Tilde{Q} = \frac{L}{\mu}$. If the step size $\alpha$ is chosen according to
\begin{equation}
\alpha = \frac{(1-\sigma^2)^2}{187\tilde{Q} L},
\end{equation}
the iterates of Algorithm \ref{alg:B} satisfy
\begin{multline}\label{eq:thm:main:rate1} 
    \frac{1}{n}\sum_{i=1}^n\mathbb{E}\|\Bar{\z}^{ST}-\z_i^{ST}\|^2+\mathbb{E}\|\Bar{\z}^{ST}-\x^\ast\|^2\leq 2\lambda^S\times\Bigg(\frac{1}{n}\sum_{i=1}^n\|\Bar{\z}^{0}-\z_i^{0}\|^2+\|\Bar{\z}^{0}-\x^\ast\|^2\\+\frac{(1-\sigma^2)^2}{1457nL^2}\sum_{i=1}^n\sum_{m=1}^d\mathbb{E}|g_{im}^{0}-\Bar{g}_m^{0}|^2\Bigg),
\end{multline}
where
\begin{equation}
    \lambda = 8\Tilde{Q}^2 \exp{(-\frac{(1-\sigma^2)^2T}{748\Tilde{Q}^2})} + 0.66,
\end{equation}
$\bar{\z}^t = \frac{1}{n}\sum_{i=1}^{2n}\z_i^t$, and $\bar{\g}^t = \frac{1}{n}\sum_{i=1}^n \g_i^t$.
\end{theorem}

\begin{corollary}\label{cor:1}
Instate the notation and hypotheses of Theorem \ref{thm:main}. If, in addition, the inner-loop duration $T$ is chosen as
\begin{equation}
 T = \mathcal{B} \lceil \frac{1496\Tilde{Q}^2}{(1-\sigma^2)^2 \mathcal{B}}\ln (200\Tilde{Q}^2) \rceil,
\end{equation}
the proposed algorithm achieves a linear convergence rate.
Furthermore, to reach an $\epsilon$-accurate solution, Algorithm \ref{alg:B} takes at most $\O(\frac{\Tilde{Q}^2\mathcal{B} \ln \Tilde{Q}}{(1-\sigma^2)^2}\ln 1/\epsilon)$ communication rounds and performs $\O((\frac{\Tilde{Q}^2\ln \Tilde{Q}}{(1-\sigma^2)^2}+ \max_i \left\{m_i \right\})\ln 1/\epsilon)$ stochastic gradient computations.
\end{corollary}
\begin{proof}
It is straightforward to verify that for the stated value of $T$, $\lambda \leq 0.7 <1$ and thus the algorithm converges linearly.
\end{proof}
Note that due to the gradient tracking step in Algorithm \ref{alg:B}, in particular when constructing the linear system of inequalities (which includes the gradient tracking error), the rate of convergence is dependent upon $\tilde{Q}^2$ (through the factor in the coefficient matrix).


\begin{remark}
Clearly, the communication cost of Algorithm \ref{alg:B} depends on the level of sparsification, i.e., the value of parameter $d_q$. Intuitively, if agents communicate fewer entries in each round, the communication cost per round decreases but the algorithm may take more rounds to reach the same accuracy. Therefore, the total communication cost until reaching a pre-specified $\epsilon$-accuracy, found as the product of the number of communication rounds and the cost per round, is of interest. Let $q$ denote the fraction of entries being communicated per iteration; smaller $q$ implies more aggressive sparsification. This compression level parameter, $q$, impacts $\sigma$ in Theorem~1; in particular, for a fixed network connectivity parameter $\mathcal{B}$, smaller $q$ leads to sparser mixing matrices and, consequently, greater $\sigma$. Note that large $\mathcal{B}$ may be caused by sparsity of the instances of a time-varying network, thus leading to large values of $\sigma$. 
\end{remark} 

\begin{remark}
It is worthwhile discussing and comparing the constants in Corollary \ref{cor:1} to those in the original SVRG \cite{johnson2013accelerating} (centralized optimization) and GT-SVRG \cite{xin2019variance} (decentralized optimization over undirected graphs). For SVRG, this constant is $O(\frac{1}{\mu \alpha(1-2L\alpha)T} + \frac{2L\alpha}{1-2L\alpha})$, where $\alpha$ denotes the step size, $L$ is the smoothness parameter, $\mu$ is the strong convexity parameter and $T$ denotes the inner loop duration \cite{johnson2013accelerating}. For both GT-SVRG and our proposed algorithm, the inner loop duration is $T = O(\frac{\Tilde{Q}^2 \log \Tilde{Q}}{(1-\sigma)^2})$ and the linear convergence constant $O(\Tilde{Q}^2 \exp{(-\frac{(1-\sigma^2)^2T}{\Tilde{Q}^2})} )$, where $\Tilde{Q}$ is the condition number and $\sigma$ is specified by the network topology and the applied compression, i.e., sparsification of the communicated quantities.

\end{remark}
\subsection{Proof of Theorem \ref{thm:main}}
In this section, we prove Theorem \ref{thm:main} by analyzing various error terms that collectively impact the convergence rate of Algorithm \ref{alg:B}. The main technical challenge in deriving the linear convergence result is the analysis of the vanishing errors formally introduced in the next paragraph: the consensus error, the optimality error and the gradient tracking error. Note that unlike in undirected graphs, the mixing matrices of directed graphs are not necessarily doubly stochastic; as a result, decentralized optimization schemes may produce state vectors that converge to a weighted average, rather than a consensus average. Furthermore, recall that in order to accelerate the convergence, we deploy two techniques: global gradient tracking and local variance reduction. Both the gradient tracking technique, which relies on the communication of gradient information over the network, and the variance reduction trick increase the difficulty of analyzing the vanishing gradient tracking error. The combination of these issues renders the analysis of the aforementioned errors challenging.

Specifically, the convergence rate depends on: (i) the expected consensus error, i.e., the expected squared difference between local vectors and the average vectors at time $(k+1) \mathcal{B}$, $\mathbb{E}[|z_{im}^{(k+1)\mathcal{B}}-\Bar{z}_{m}^{(k+1)\mathcal{B}} |^2]$; (ii) the expected optimality error, i.e., the expected squared difference between the average vectors and the optimal vector, $\mathbb{E}[\|\Bar{\z}^{(k+1)\mathcal{B}} - \x^* \|^2]$; and (iii) the expected gradient tracking error, $\mathbb{E}[\sum_{m=1}^d \sum_{i=1}^n |g_{im}^{(k+1)\mathcal{B}}-\Bar{g}_m^{(k+1)\mathcal{B}} |^2$. Hence, it is critical to determine evolution of these sequences. Note that compared to the gradient-tracking based work, e.g. \cite{nedic2017achieving, xin2019variance}, analyzing the proposed scheme is more involved due to its reliance upon a combination of variance reduction techniques and a communication-sparsified consensus; showing that the novel scheme achieves linear convergence on general directed time-varying graphs despite sparsified communication calls for a careful examination of the error terms in a manner distinct from the analysis found in prior work.

Dynamics of the aforementioned errors are clearly interconnected. Consequently, our analysis relies on deriving recursive bounds for the errors in terms of the linear combinations of their past values. The results are formally stated in Lemma \ref{lem:consensus_error}, Lemma \ref{lem:optimality_error}, and Lemma \ref{lem:gt_error}. Proofs of these lemmas are provided in the supplementary document.

\begin{lemma}\label{lem:consensus_error}
Suppose Assumption \ref{new_assp} holds. Then $\forall i \leq n$, $k \geq 0$ and $0 < m \leq d$, updates generated by Algorithm \ref{alg:B} satisfy
\begin{equation}
\begin{aligned}
    \mathbb{E}[|z_{im}^{(k+1)\mathcal{B}}-\Bar{z}_{m}^{(k+1)\mathcal{B}} |^2] & \leq \frac{1+\sigma^2}{2}\mathbb{E}[|z_{im}^{k\mathcal{B}}-\Bar{z}_{m}^{k\mathcal{B}} |^2 ]  +\frac{2\alpha^2}{1-\sigma^2}\mathbb{E}[|g_{im}^{k\mathcal{B}}-\Bar{g }_{m}^{k\mathcal{B}} |^2 ].
\end{aligned}
\end{equation}
\end{lemma}
Having established in Lemma \ref{lem:consensus_error} a recursive bound on the expected consensus error, we proceed by stating in Lemmas \ref{lem:optimality_error} and \ref{lem:gt_error} recursive bounds on the expected optimality and gradient tracking errors, respectively. First, let us introduce (for $k \geq 0$)
\begin{equation}
\begin{aligned}
    \Bar{\tau}^{k\mathcal{B}} &= \frac{1}{n}\sum_{i=1}^n \tau_i^{k\mathcal{B}},\\
    \tau_i^{(k+1)\mathcal{B}} &= \begin{cases}
    \x_i^{(k+1)\mathcal{B}} & \mathrm{if} \quad (k+1)\mathcal{B} \mod T = 0 \\
    \tilde{w}_i & \mathrm{otherwise}.
    \end{cases}
\end{aligned}
\end{equation}
\begin{lemma}\label{lem:optimality_error}
Suppose Assumption \ref{new_assp} holds and let $0 < \alpha < \frac{\mu}{8L^2}$. Then for all $k>0$ it holds that
\begin{align}
    \begin{split}
    \mathbb{E}[n \|\Bar{\z}^{(k+1)\mathcal{B}} - \x^* \|^2] & \leq  \frac{2L^2 \alpha}{\mu}\mathbb{E}[\sum_{i=1}^n  \|\Bar{\z}^{k\mathcal{B}}-\z_i^{k\mathcal{B}} \|^2 ] \\ & \quad + (1-\frac{ \mu \alpha}{2} )\mathbb{E}[n\|\Bar{\z}^{k\mathcal{B}}-\x^* \|^2 ] \\
    & \quad + \frac{4L^2 \alpha^2}{n} \mathbb{E}[ \sum_{i=1}^n \| \tau_i^{k\mathcal{B}}-\Bar{\tau}^{k\mathcal{B}} \|^2] \\ & \quad + \frac{4L^2 \alpha^2}{n} \mathbb{E}[n \|\Bar{\tau}^{k\mathcal{B}} - \x^* \|^2 ].
\end{split}
\end{align}
\end{lemma}
\begin{lemma}\label{lem:gt_error}
Suppose Assumption \ref{new_assp} holds. Then
\begin{equation}
\begin{aligned}
\frac{1 }{L^2}\mathbb{E}[\sum_{m=1}^d \sum_{i=1}^n |g_{im}^{(k+1)\mathcal{B}}-\Bar{g}_m^{(k+1)\mathcal{B}} |^2]   & \leq \frac{120}{1-\sigma^2}\mathbb{E}[\sum_{i=1}^n  \|\Bar{\z}^{k\mathcal{B}}-\z_i^{k\mathcal{B}} \|^2  ]\\& \quad + \frac{89}{1-\sigma^2}\mathbb{E}[ n\|\Bar{\z}^{k\mathcal{B}}-\x^* \|^2] \\& \quad + \frac{3+\sigma^2}{4} \mathbb{E}[\frac{\sum_{m=1}^d \sum_{i=1}^n |g_{im}^{k\mathcal{B}}-\Bar{g}_m^{k\mathcal{B}} |^2  }{L^2} ]\\ & \quad + \frac{38}{1-\sigma^2} \mathbb{E}[\sum_{i=1}^n  \|\Bar{\z}^{k\mathcal{B}}-\z_i^{k\mathcal{B}} \|^2  ] \\ & \quad
+ \frac{38}{1-\sigma^2} \mathbb{E}[n \|\Bar{\tau}^{k\mathcal{B}} - \x^* \|^2 ].
\end{aligned}
\end{equation}
\end{lemma}
We proceed by defining a system of linear inequalities  involving the three previously discussed error terms; study of the conditions for the geometric convergence of the powers of the resultant matrix in the system of linear inequalities leads to the linear convergence result in Theorem \ref{thm:main}. To this end, we first state Proposition \ref{proposition:LTI} whose proof follows by combining and re-organizing the inequalities in Lemmas \ref{lem:consensus_error}-\ref{lem:gt_error} in a matrix form.
\begin{proposition}\label{proposition:LTI}
Suppose Assumption \ref{new_assp} holds. Define 
\begin{equation}
    \u^{k\mathcal{B}} = \begin{bmatrix}
    \mathbb{E}[\sum_{i=1}^n  \|\Bar{\z}^{k\mathcal{B}}-\z_i^{k\mathcal{B}} \|^2  ] \\
    \mathbb{E}[ n\|\Bar{\z}^{k\mathcal{B}}-\x^* \|^2] \\
    \mathbb{E}[\frac{\sum_{m=1}^d \sum_{i=1}^n |g_{im}^{k\mathcal{B}}-\Bar{g}_m^{k\mathcal{B}} |^2  }{L^2} ]
    \end{bmatrix},
\end{equation}

\begin{equation}
    \Tilde{\u}^{k\mathcal{B}} = \begin{bmatrix}
     \mathbb{E}[\sum_{i=1}^n \| \tau_i^{k\mathcal{B}}-\Bar{\tau}^{k\mathcal{B}} \|^2 ] \\
     \mathbb{E}[n \|\Bar{\tau}^{k\mathcal{B}} - \x^* \|^2 ] \\
     \mathbf{0}
    \end{bmatrix},
\end{equation}

\begin{equation}
    J_{\alpha} = \begin{bmatrix}
    \frac{1+\sigma^2}{2} & 0 & \frac{2\alpha^2L^2}{1-\sigma^2} \\
    \frac{2L^2\alpha}{\mu} & 1-\frac{\mu \alpha}{2} & 0 \\
    \frac{120}{1-\sigma^2} & \frac{89}{1-\sigma^2} & \frac{3+\sigma^2}{4} 
    \end{bmatrix},
\end{equation}

\begin{equation}
    H_{\alpha} = \begin{bmatrix}
    0 & 0 & 0 \\
    \frac{4L^2 \alpha^2}{n} & \frac{4L^2 \alpha^2}{n} & 0 \\
    \frac{38}{1-\sigma^2} & \frac{38}{1-\sigma^2} & 0
    \end{bmatrix}.
\end{equation}
If $0 \leq \alpha \leq \frac{\mu(1-\sigma^2)}{14\sqrt{2}L^2}$, then for any
$k \geq 0$ it holds that
\begin{equation}
    \u^{(k+1)\mathcal{B}} \leq J_{\alpha}\u^{k\mathcal{B}} + H_{\alpha} \Tilde{\u}^{k\mathcal{B}}.
\end{equation}
\end{proposition}
As a direct consequence of Proposition \ref{proposition:LTI}, for the iterations of the inner loop of Algorithm \ref{alg:B}, for all $ k \in [s\lfloor T/ \mathcal{B} \rfloor, (s+1)\lfloor T/ \mathcal{B} \rfloor-1]$ it holds that
\begin{equation}
    \u^{(k+1)\mathcal{B}} \leq J_{\alpha}\u^{k\mathcal{B}} + H_{\alpha} \u^{sT},
\end{equation}
while for the outer loop of Algorithm \ref{alg:B} it holds for all $\forall s \geq 0$,
\begin{equation}\label{eq:outer_loop_LTI}
    \u^{(s+1)T} \leq (J_{\alpha}^T + \sum_{l=0}^{T-1}J_{\alpha}^l H_{\alpha}) \u^{sT}.
\end{equation}

Now, to guarantee linear decay of the outer loop sequence, we restrict the range of the inner loop duration $T$ and the step size $\alpha$ according to
\begin{equation}
    \rho(J_{\alpha}^T + \sum_{l=0}^{T-1}J_{\alpha}^l H_{\alpha} ) < 1,
\end{equation}
where $\rho(\cdot)$ denotes the spectral radius of its argument.

In Lemma \ref{lem:LTI_convergence} below, we establish the range of $\alpha$ such that the weighted matrix norms of $J_{\alpha}^T$ and $ \sum_{l=0}^{T-1}J_{\alpha}^l H_{\alpha}$ are small, thereby ensuring the geometric convergence of the powers of these matrices to $\mathbf{0}$.
\begin{lemma}\label{lem:LTI_convergence} 
Suppose Assumption \ref{new_assp} holds and let $ 0 < \alpha \leq \frac{(1-\sigma^2)^2}{187\Tilde{Q}L}$, where $\Tilde{Q} = \frac{L}{\mu}$.
Then
\begin{equation}
    \rho(J_{\alpha}) < \| |J_{\alpha}| \|^{\mathbf{\delta}}_{\infty} < 1 - \frac{\mu \alpha}{4}
\end{equation}
and
\begin{equation}
    \| |\sum_{l=0}^{T-1}J_{\alpha}^l H_{\alpha}| \|^{\q}_{\infty} \leq
     \| | (I - J_{\alpha})^{-1} H_{\alpha} | \|^{\q}_{\infty}< 0.66,
\end{equation}
where $\mathbf{\delta} = \begin{bmatrix}
1, 8\Tilde{Q}^2, \frac{6656\Tilde{Q}^2}{(1-\sigma^2)^2}
\end{bmatrix}$ and $\q = [1, 1, \frac{1457}{(1-\sigma^2)^2}]$. 
\end{lemma}



Essentially, Lemma 5 establishes the range of the stepsize $\alpha$ such that the matrices involved in the system of linear inequalities \eqref{eq:outer_loop_LTI} have small norm. Upon setting $\alpha = \frac{(1-\sigma^2)^2}{187\Tilde{Q}L}$ (i.e., assigning $\alpha$ the largest feasible value), we proceed to determine the number of iterations in the outer loop such that the powers of matrix $J_{\alpha}^T + \sum_{l=0}^{T-1}J_{\alpha}^l H_{\alpha}$ in \eqref{eq:outer_loop_LTI}, and hence the components of $\u$ (i.e. the error terms), converge to zero at a geometric rate.


To this end, note that since $J_{\alpha}$ is non-negative,
$\sum_{l=0}^{T-1}J_{\alpha}^l \leq \sum_{l=0}^{\infty}J_{\alpha}^l = (I - J_{\alpha})^{-1} $ .
Hence, for all $ s \geq 0$ it holds 
\begin{equation}
    \u^{(s+1)T} \leq (J_{\alpha}^T + (I - J_{\alpha})^{-1} H_{\alpha}) \u^{sT}.
\end{equation}
Since $\alpha = \frac{(1-\sigma^2)^2}{187QL}$, we may write
\begin{align}
\begin{split}
    \|\u^{(s+1)T}\|_{\infty}^{\q} & \leq \||J_{\alpha}^T + (I - J_{\alpha})^{-1} H_{\alpha}| \|_{\infty}^{\q}  \|\u^{sT}\| _{\infty}^{\q}  \\
    & \leq ( \||J_{\alpha}^T | \|_{\infty}^{\q} +0.66) \|\u^{sT}\| _{\infty}^{\q} \\
    & \leq (8\Tilde{Q}^2 (\||J_{\alpha}^T | \|_{\infty}^{\delta} )^T + 0.66) \|\u^{sT}\| _{\infty}^{\q} \\
    & \leq (8\Tilde{Q}^2 \exp{(-\frac{(1-\sigma^2)^2T}{748\Tilde{Q}^2})} + 0.66)\|\u^{sT}\| _{\infty}^{\q} \\
    &:=\lambda \|\u^{sT}\| _{\infty}^{\q}.
\end{split}
\end{align}
The result in \eqref{eq:thm:main:rate1} follows simply by noting the definitions of $\u^{sT}$ and the $\| \cdot \|^\q_\infty$ norm. Therefore, the proof of Theorem \ref{thm:main} is complete.

\section{Experiments}\label{sec:exp}

In this section, we report results of benchmarking the proposed Algorithm \ref{alg:B}; for convenience, we refer to 
Algorithm \ref{alg:B} as Di-CS-SVRG (\underline{Di}rected \underline{C}ommunication-\underline{S}parsified \underline{S}tochastic \underline{V}ariance-\underline{R}educed \underline{G}radient Descent). The results for the proposed Algorithm \ref{alg:A} are presented in the supplementary material. 

We start with a network consisting of $10$ nodes with randomly generated time-varying connections while ensuring strong connectivity at each time step. The construction begins with the Erdős–Rényi model \cite{erdos1959random} where each edge exists independently with probability $0.9$; then, $2$ directed edges are dropped from each strongly connected graph, leading to directed graphs. Building upon this basic structure, we can design networks with different connectivity profiles. Recall that the window size parameter $\mathcal{B}$, introduced in Assumption \ref{new_assp}(\ref{assumption5}),
may imply that the union graph over $\mathcal{B}$ consecutive instances, starting from any instance that is a multiple of $\mathcal{B}$, forms an almost-surely strongly connected Erdős–Rényi graph. When $\mathcal{B} = 1$, the network is strongly connected at each time step. The parameter $q$, the fraction of entries being communicated to neighboring nodes, characterizes the level of message sparsification; $q = 1$ implies communication without compression, while $q = 0$ indicates there is no communication in the network.

\subsection{Decentralized Optimization Problem}\label{sec42}


We test the proposed Di-CS-SVRG on two tasks, linear and logistic regression, and benchmark it against several baseline algorithms. In particular, we compare Di-CS-SVRG with Decentralized Stochastic Gradient Descent (De-Stoc, stochastic variant), Push-DIGing (Push-DIG-Full) \cite{nedic2017achieving}, Push-DIGing Stochastic (Push-DIG-Stoc, stochastic variant) and the TV-AB algorithm (AB Algorithm) \cite{saadatniaki2018optimization}. In addition, we include comparisons to the full gradient schemes we considered in our preliminary work \cite{chen2020communication},
Decentralized Full Gradient Descent (De-Full) and its communication-sparsified variant Sparsified De-Full (S-De-Full).

\subsubsection{Decentralized Linear Regression} We consider the setting where $n$ nodes collaboratively solve the optimization problem
\begin{equation}
    \min_{\x}\left\{  \frac{1}{n}\sum_{i=1}^n \|\mathbf{y}_i-D_i \mathbf{x}\|^2\right\},
\end{equation}
where for the $i^{th}$ node $D_i \in \mathbb{R}^{200 \times 64}$ denotes the matrix with $200$ local samples of size $d=64$, and $\mathbf{y}_i \in \mathbb{R}^{200}$ denotes the corresponding measurement vector. The true value of $\mathbf{x}$, $\mathbf{x}^*$, is generated from a normal distribution,
and the samples are generated synthetically. The measurements are generated as $\mathbf{y}_i=M_i \mathbf{x}^*+\eta_i$, where the entries of $M_i$ are generated randomly from the standard normal distribution and then $M_i$ is normalized such that its rows sum up to one. The local noise vector $\eta_i$ is drawn at random from a zero-mean Gaussian distribution with variance $0.01$. All algorithms are initialized using randomly generated local vectors $\mathbf{x}_i^0$, 
and utilize constant step size $\alpha_t=0.002$ except the AB algorithm for which we followed the recommendation in \cite{saadatniaki2018optimization} and set $\alpha_t = 0.0025$.

\begin{figure*}
        \centering
        \begin{subfigure}[b]{0.475\textwidth}
            \centering
            \includegraphics[width=\textwidth]{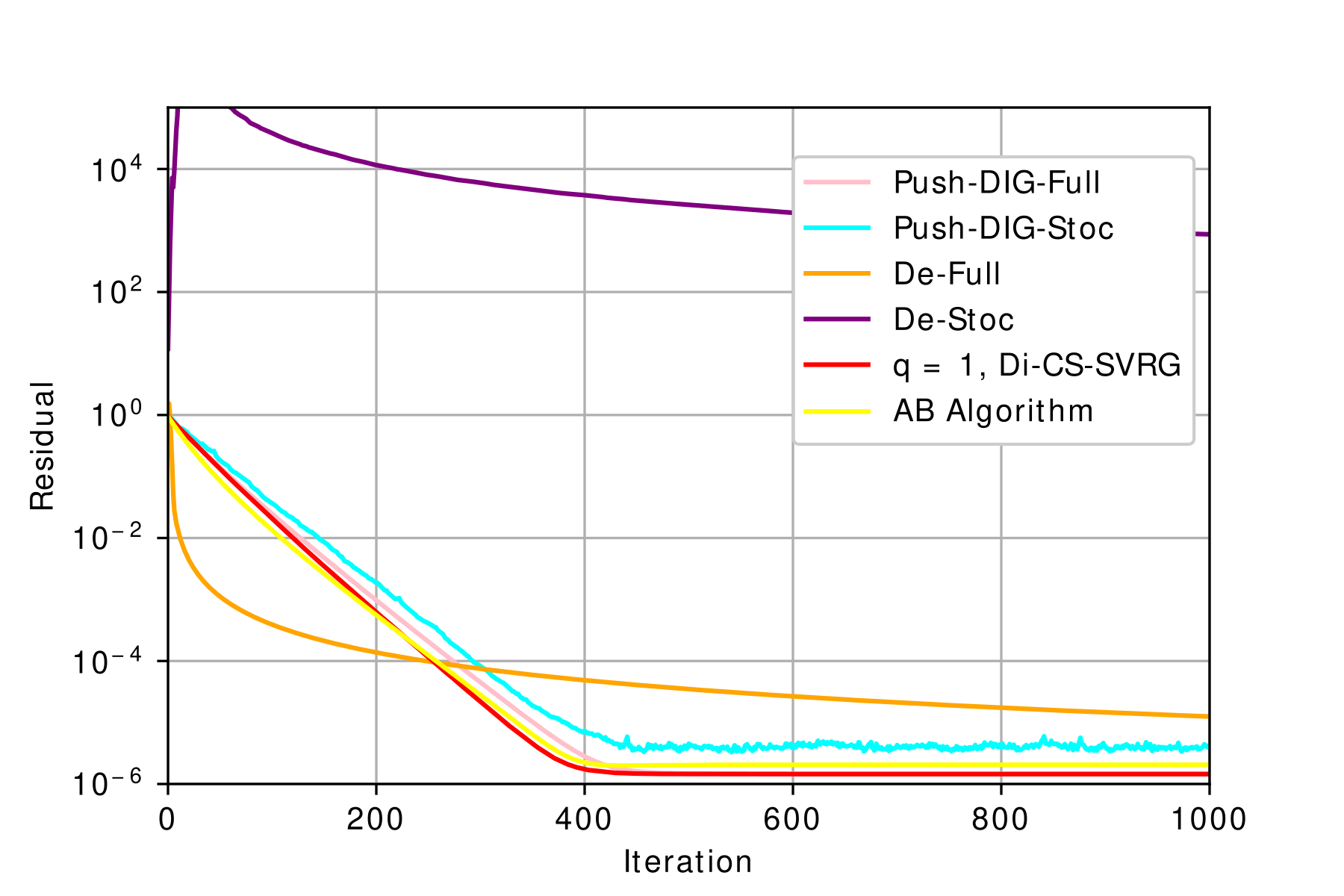}
            \caption[Network2]%
            {{ Residual vs. iterations: full communication schemes.}}    
            \label{fig:mean and std of net14}
        \end{subfigure}
        \hfill
        \begin{subfigure}[b]{0.475\textwidth}  
            \centering 
            \includegraphics[width=\textwidth]{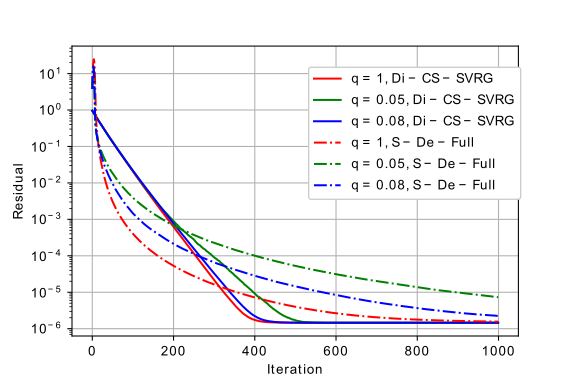}
            \caption[]%
            {{\small Residual vs. iterations: compressed communication schemes.}}    
            \label{fig:mean and std of net24}
        \end{subfigure}
        \vskip\baselineskip
        \begin{subfigure}[b]{0.475\textwidth}   
            \centering 
            \includegraphics[width=\textwidth]{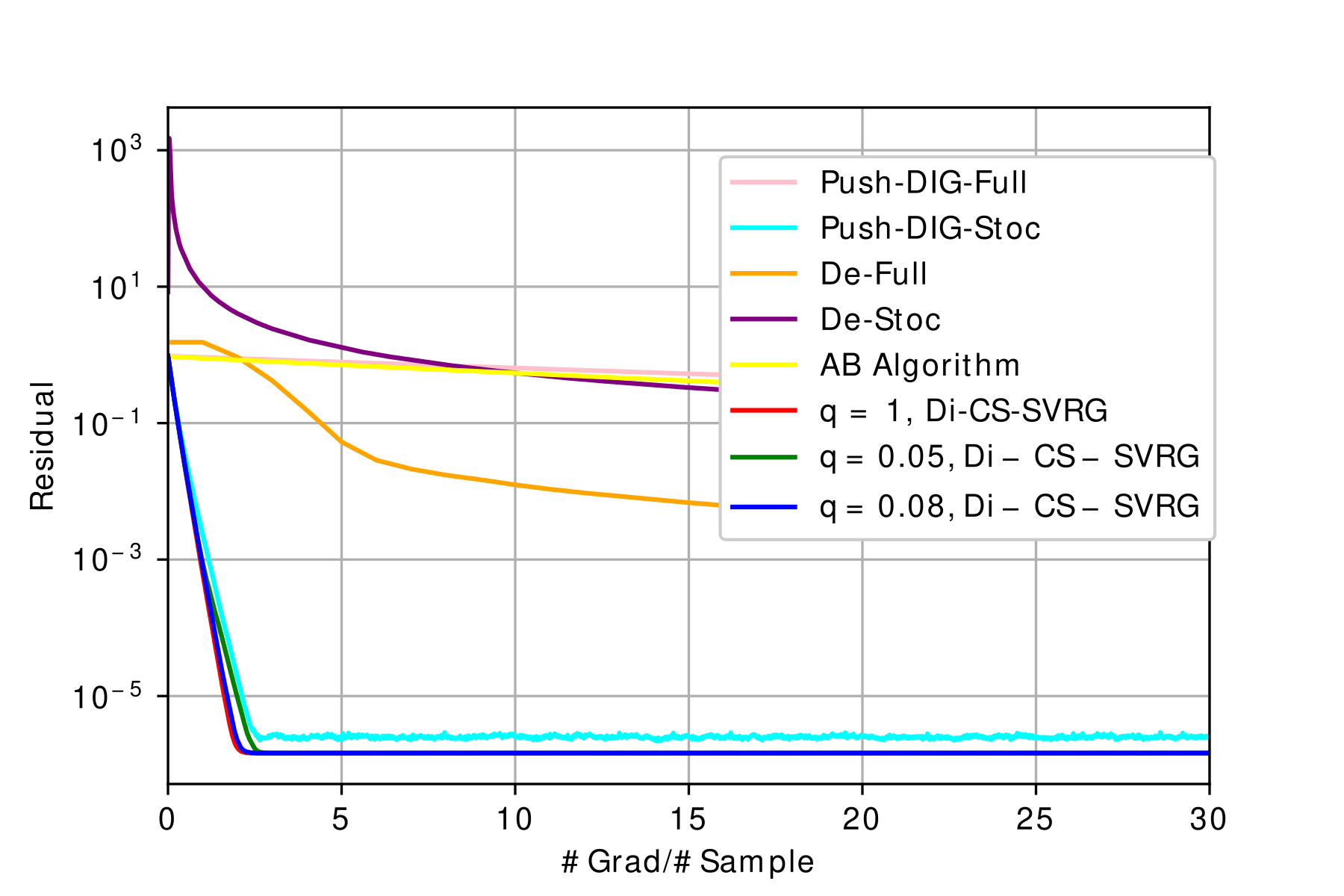}
            \caption[]%
            {{\small Residual vs. gradient computations per sample.}}    
            \label{fig:mean and std of net34}
        \end{subfigure}
        \hfill
        \begin{subfigure}[b]{0.475\textwidth}   
            \centering 
            \includegraphics[width=\textwidth]{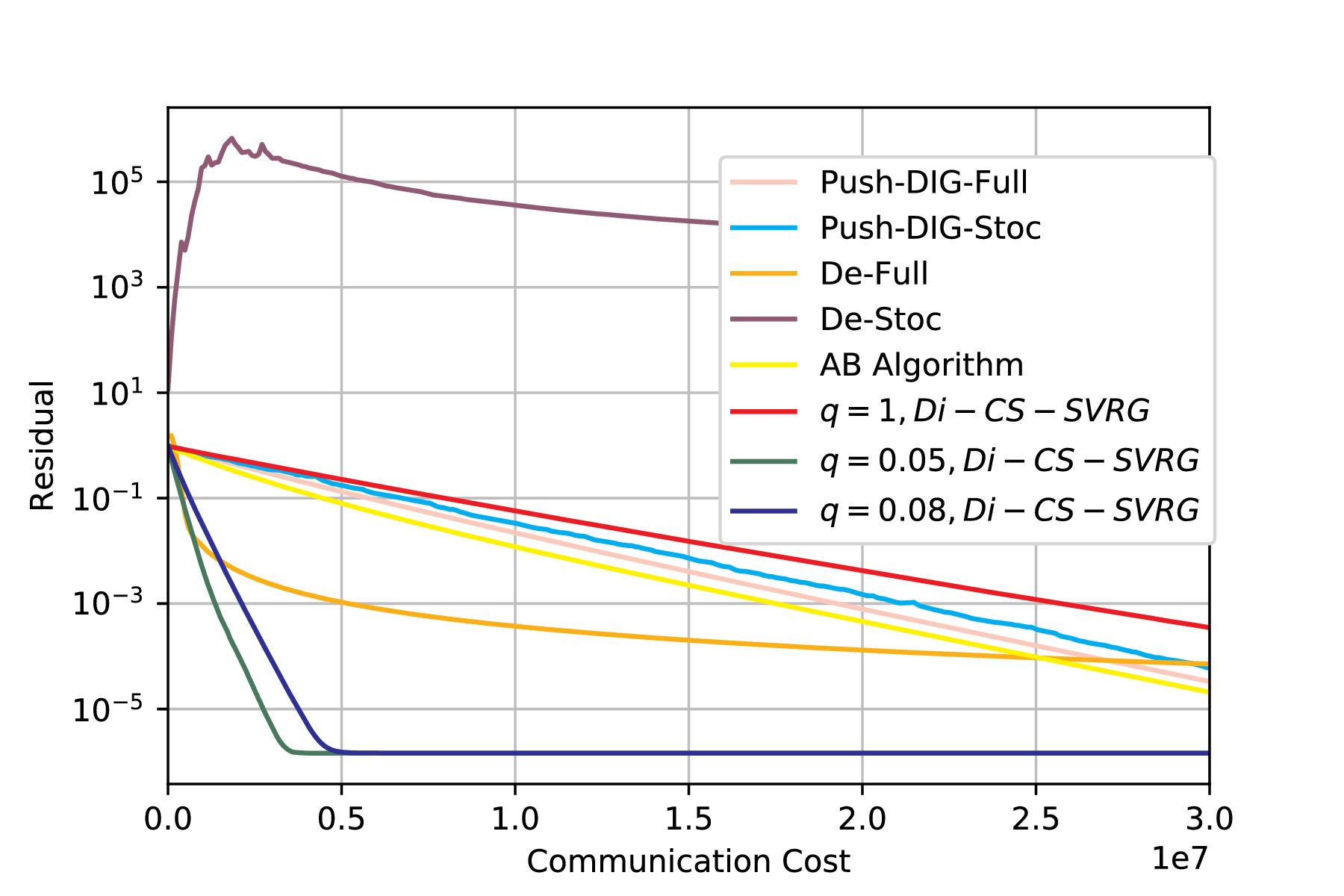}
            \caption[]%
            {{\small Residual vs. communication cost.}}    
            \label{fig:mean and std of net44}
        \end{subfigure}
        \caption[ The average and standard deviation of critical parameters ]
        {\small Linear regression, $\mathcal{B} = 5$. (a) The residual achieved by full communication schemes and the residual of Di-CS-SVRG (Algorithm \ref{alg:B}) with $q = 1$ vs. iterations. (b) The residual achieved by Di-CS-SVRG with different compression levels: $q = 1$, $q=0.08$, and $q=0.05$ vs. iterations. (c) The cumulative number of gradient computations needed to reach given level of the residual; showing both the compressed as well as full communication schemes. (d) The cumulative communication cost to reach given level of the residual; showing both the compressed as well as full communication schemes.} 
        \label{fig:lin_algo1}
    \end{figure*}

Performance of the algorithms is characterized using three metrics: residual over iterations, residual over average gradient computation and residual over communication cost, where the residual is computed as $\frac{\|\mathbf{x}^t-\mathbf{x}^*\|}{\|\mathbf{x}^0-\mathbf{x}^*\|}$. The results are shown in Fig. \ref{fig:lin_algo1}. As seen in Fig. \ref{fig:lin_algo1}(a),  Di-CS-SVRG (i.e., our Algorithm \ref{alg:B}) with $q=1$ converges at a linear rate and, while being a stochastic gradient algorithm, reaches the same residual floor as the full gradient method Push-DIG-Full and the AB algorithm. Di-CS-SVRG converges much faster than the two baseline algorithms employing SGD, Push-DIG-Stoc and De-Stoc. Fig. \ref{fig:lin_algo1}(b) shows that Di-CS-SVRG with varied compression levels $q$ converges to the same residual floor, and that (as expected) larger $q$ leads to faster convergence. Moreover, the figure shows that for a fixed $q$, Di-CS-SVRG achieves faster convergence than the benchmark algorithms. 

Fig. \ref{fig:lin_algo1}(c) compares different algorithms in terms of the number of gradients computed per sample, demonstrating computation efficiency of Di-CS-SVRG. Finally, Fig. \ref{fig:lin_algo1}(d) shows the communication cost for varied $q$, computed as the total number of the (state, surplus and gradient) vector entries communicated across the network. As seen in the figure, to achieve a pre-specified level of the residual, Di-CS-SVRG with $q = 0.05$ incurs smaller communication cost than any other considered algorithm. 

\begin{figure*}
        \centering
        \begin{subfigure}[b]{0.475\textwidth}
            \centering
            \includegraphics[width=\textwidth]{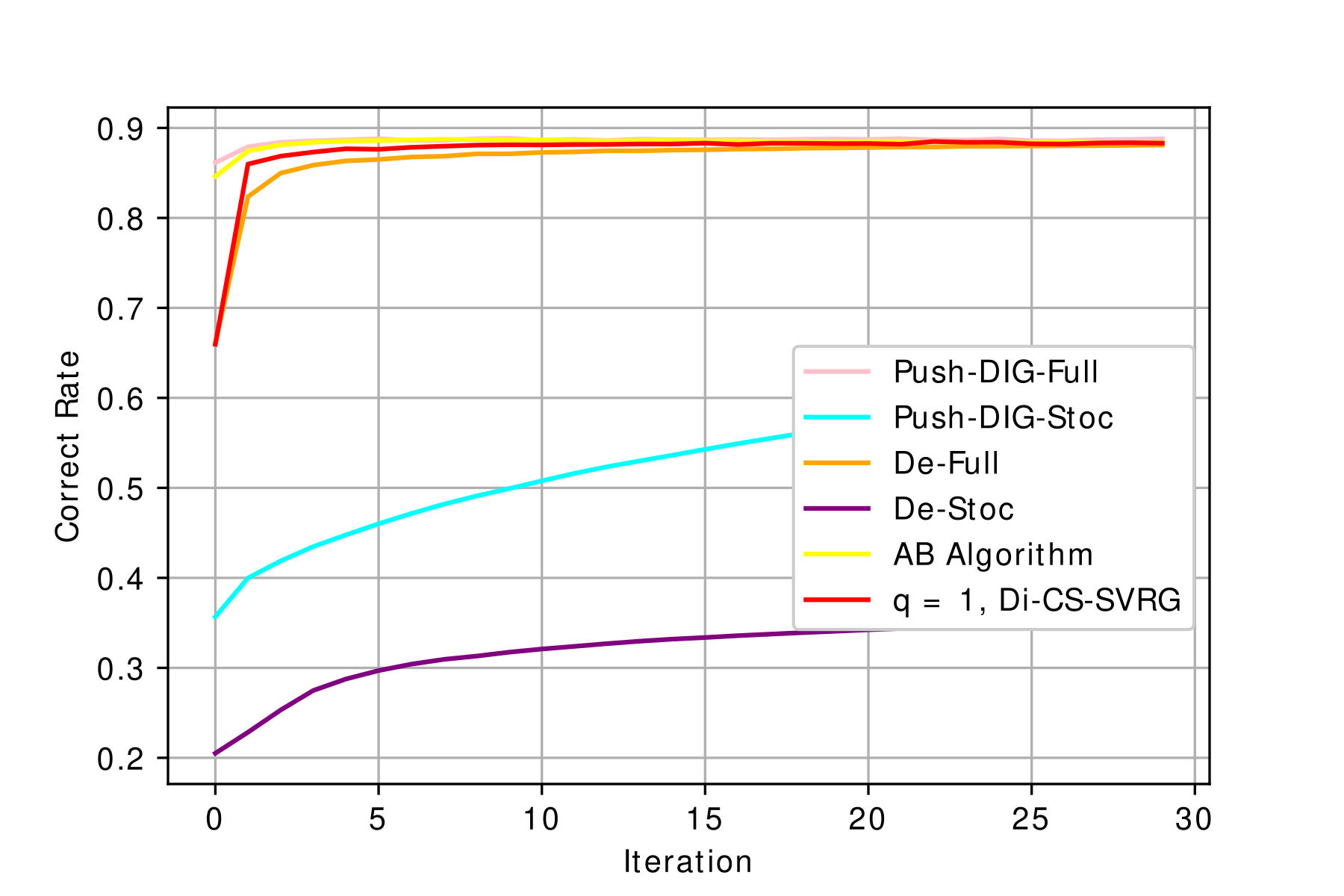}
            \caption[Network2]%
            {{\small Correct rate vs. iterations: full communication schemes.}}    
            \label{fig:mean and std of net141}
        \end{subfigure}
        \hfill
        \begin{subfigure}[b]{0.475\textwidth}  
            \centering 
            \includegraphics[width=\textwidth]{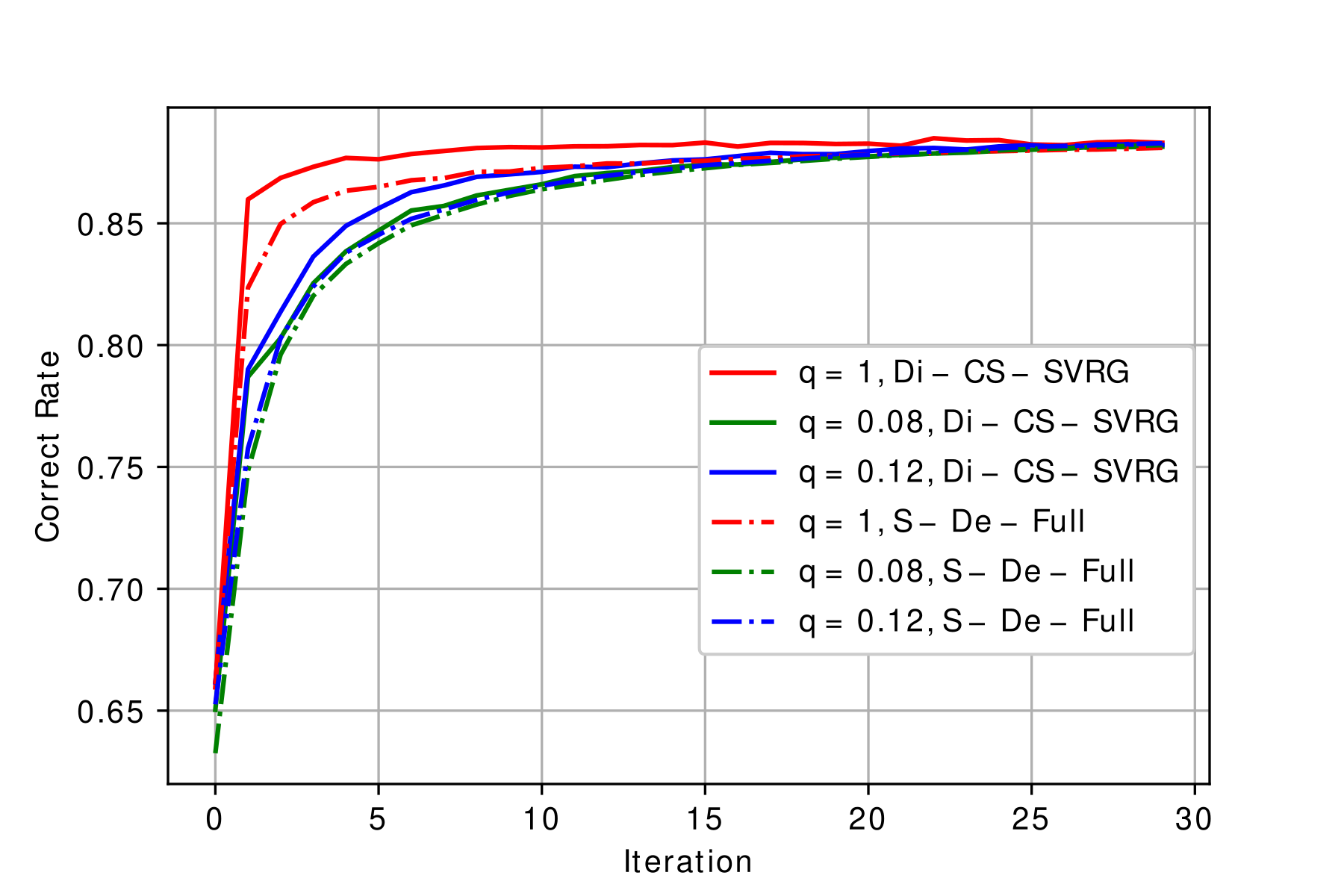}
            \caption[]%
            {{\small Correct rate vs. iterations: compressed schemes.}}    
            \label{fig:mean and std of net241}
        \end{subfigure}
        \vskip\baselineskip
        \begin{subfigure}[b]{0.475\textwidth}   
            \centering 
            \includegraphics[width=\textwidth]{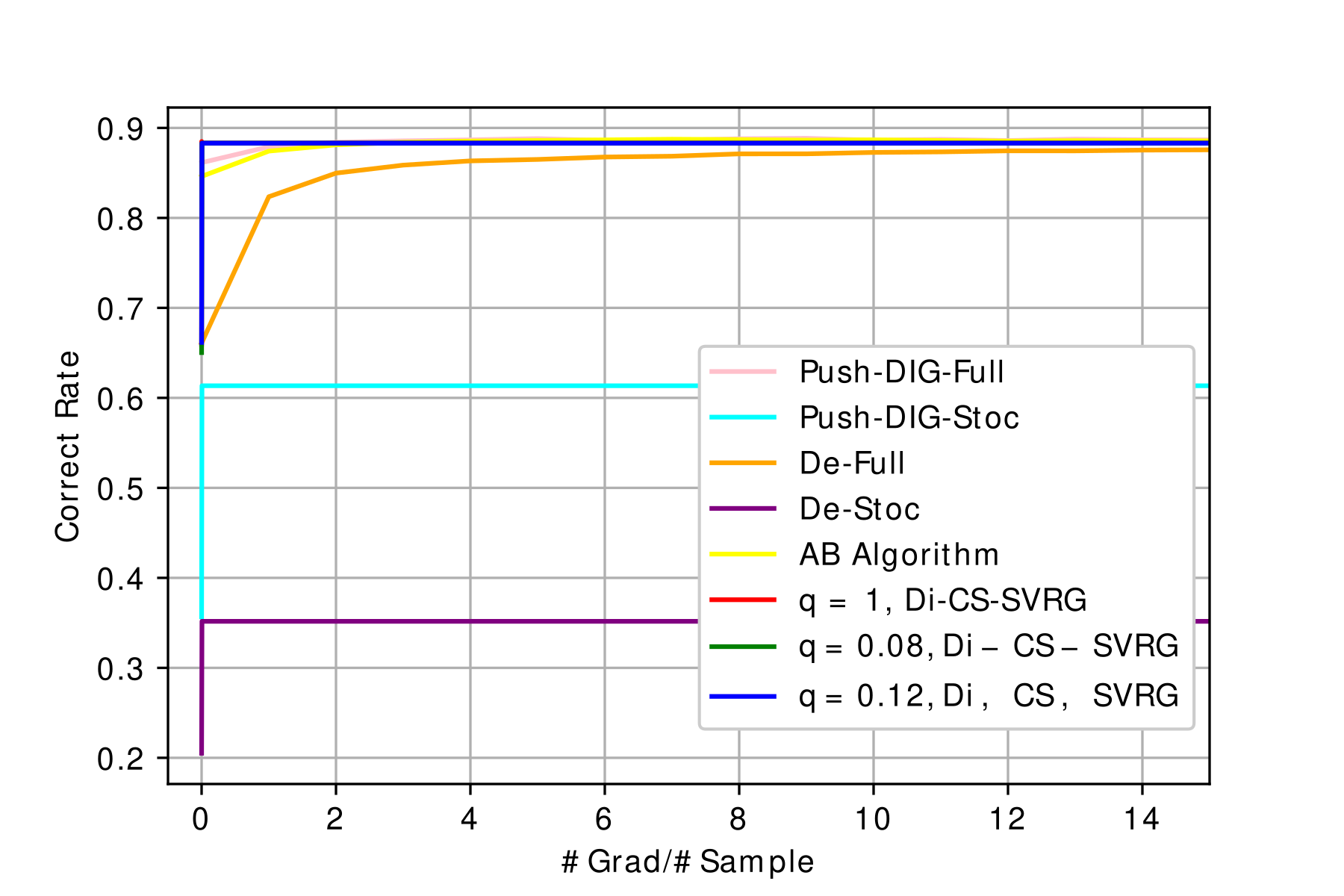}
            \caption[]%
            {{\footnotesize Correct rate vs. gradient computations per sample.}}    
            \label{fig:mean and std of net341}
        \end{subfigure}
        \hfill
        \begin{subfigure}[b]{0.475\textwidth}   
            \centering 
            \includegraphics[width=\textwidth]{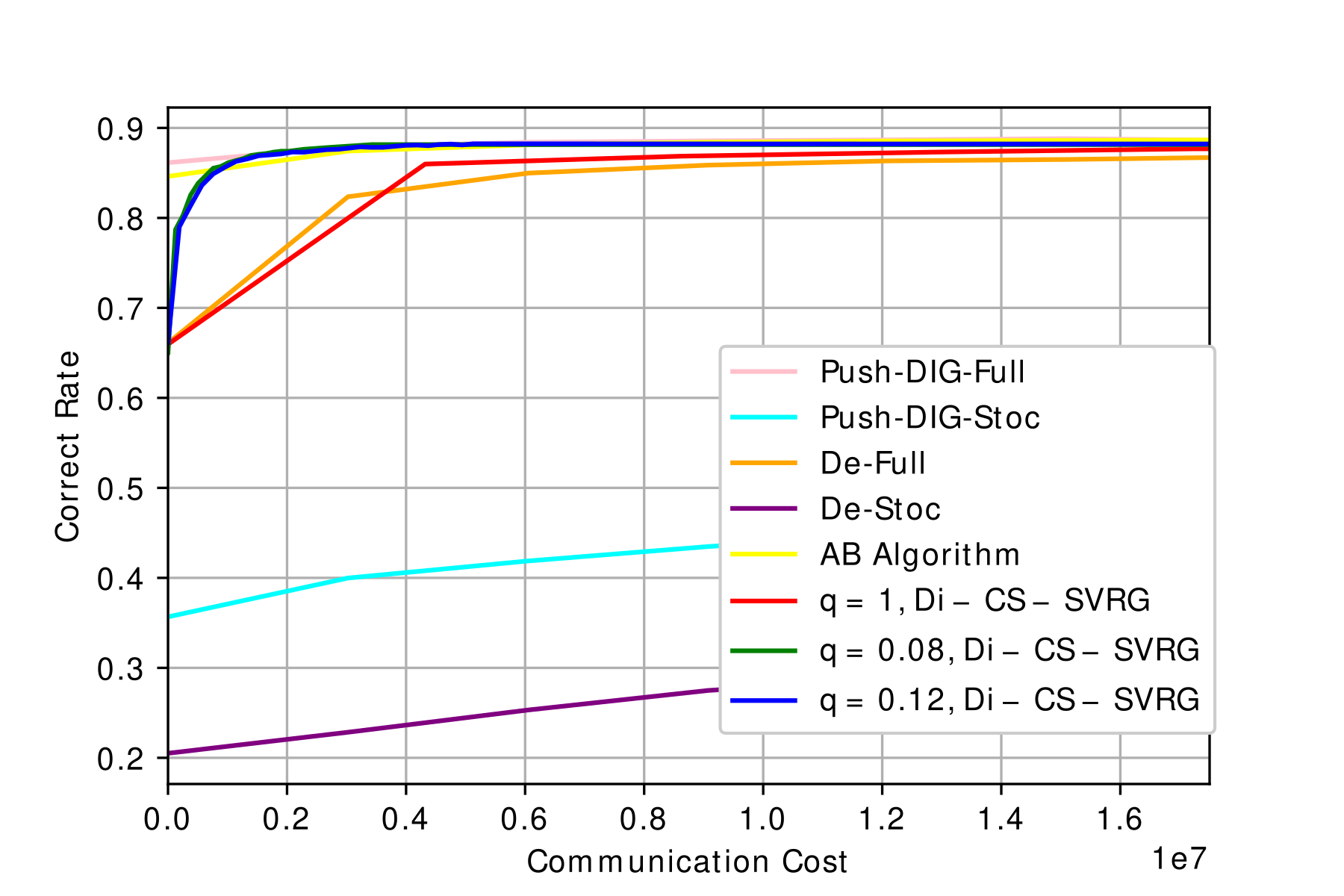}
            \caption[]%
            {{\small Correct rate vs. communication cost.}}   
            \label{fig:mean and std of net441}
        \end{subfigure}
        \caption[ The average and standard deviation of critical parameters ]
        {\small Logistic regression, $\mathcal{B} = 1$. (a) The correct classification rate achieved by full communication schemes and the correct classification rate of  Di-CS-SVRG vs. iterations. (b) The correct classification rate for Di-CS-SVRG with varied compression levels: $q = 1$, $q=0.12$, $q=0.08$ vs. iterations.
	(c) The cumulative number of gradient computations required to reach given levels of correct classification rate. (d) The cumulative communication cost required to reach given level of the correct classification rate.} 
        \label{fig:logi}
    \end{figure*}


\subsubsection{Decentralized Logistic Regression} To perform benchmarking on a logistic regression task, we solve a multi-class classification problem on the Stackoverflow dataset \cite{kaggle}.
\begin{equation}
\min_{\x} \left\{ \frac{\mu}{2}\|\mathbf{x}\|^2+\sum_{i=1}^n \sum_{j=1}^N \mathrm{ln}(1+\mathrm{exp}(-(\mathbf{m}_{ij}^T\mathbf{x})\y_{ij})) \right\},
\end{equation}
where the training samples $(\mathbf{m}_{ij}, \y_{ij}) \in \mathbb{R}^{400+5}$,  $\mathbf{m}_{ij}$ represents a vectorized text feature and $\y_{ij}$ represents the corresponding tag vector. We compare the performance of Di-CS-SVRG with the same benchmarking algorithms as in the linear regression problem, and use the same initialization setup. 
The logistic regression experiment is run with the stepsize $\alpha_t=0.01$; the regularization parameter is set to $\mu=10^{-5}$. 

Performance of the algorithms on the logistic regression task is characterized by the classification correct rate. In particular, we evaluate the following three metrics:
the correct rate vs. iterations, the correct rate vs. average gradient computation, and the correct rate vs. communication cost; they are all shown in Fig. \ref{fig:logi}. As seen in Fig. \ref{fig:logi}(a),  Di-CS-SVRG converges and reaches the same residual floor as the full gradient method Push-DIG-Full and the AB Algorithm. Di-CS-SVRG converges much faster than then the two algorithms that rely on SGD, Push-DIG-Stoc and De-Stoc. Fig. \ref{fig:logi}(b) shows that for varied compression levels $q$, Di-CS-SVRG converges to the same residual floor. As expected, larger $q$ leads to faster convergence. For a fixed $q$, Di-CS-SVRG converges faster than the benchmark algorithm. 

Fig. \ref{fig:logi}(c) reports the average gradient computation, i.e., the number of gradients computed per sample. As can be seen, Di-CS-SVRG with different compression levels uses fewer gradient computation than the full gradient schemes (Push-DIG-Full, the AB algorithm and De-Full) to reach $90\%$ correct classification rate.

Fig. \ref{fig:logi}(d) shows the communication cost, defined as the total number of the (state, surplus and gradient) vector entries exchanged across the network, for various values of $q$. Among the considered schemes, Di-CS-SVRG with $q=0.08$ reaches the pre-specified residual level with smaller communication cost than any stochastic scheme (Push-DIG-Stoc, De-Stoc as well as Di-CS-SVRG with other $q$'s). Even though Push-DIG-Stoc reaches various accuracies slightly faster than Di-CS-SVRG, it requires considerably higher amount of gradient computation to do so (see Fig. \ref{fig:logi}).

\subsection{Results on different network topologies}

To further test Di-CS-SVRG in different settings, we apply it to decentralized optimization over networks with varied connectivity and sizes.

\subsubsection{Varied network connectivity}
\begin{figure}[htbp]
\subfloat[Linear regression for varied network connectivity.]{%
  \includegraphics[clip,width=0.9\columnwidth]{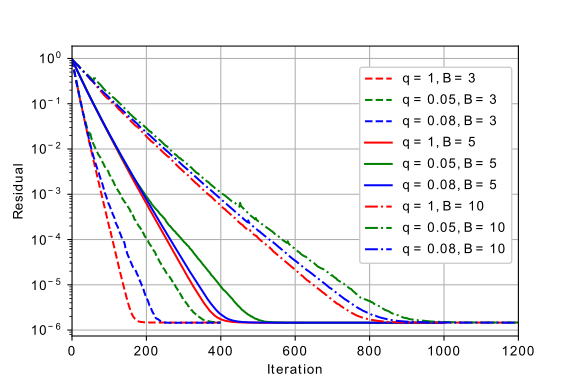}%
}

\subfloat[Correct rate on logistic regression with varied network sizes.]{%
  \includegraphics[clip,width=0.9\columnwidth]{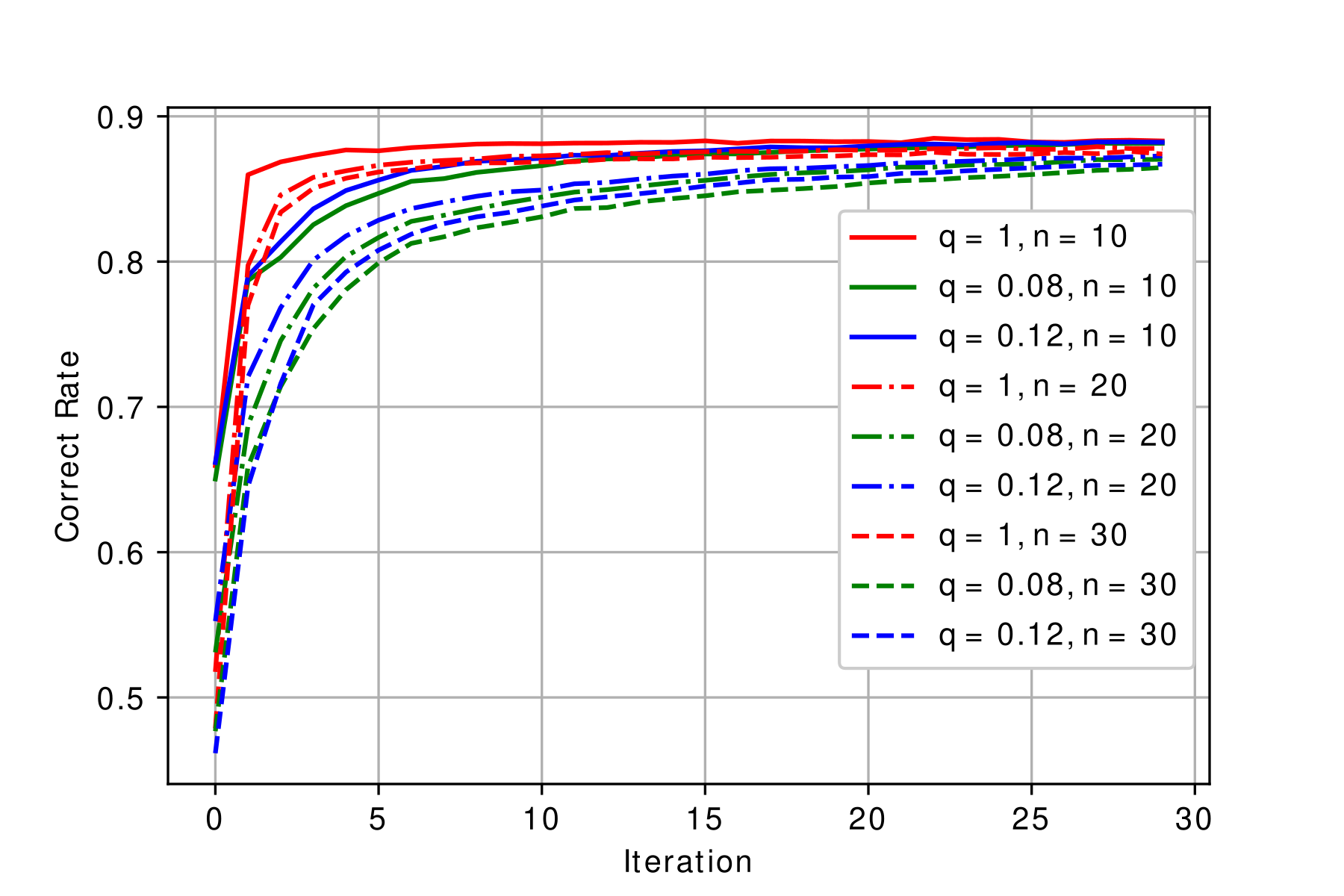}%
}
\caption{Varied network connectivity and size.}
\label{fig:expe_more}
\end{figure}

We consider the linear regression problem and vary the values of the joint connectivity parameter $\mathcal{B}$. Fig. \ref{fig:expe_more}(a) shows the resulting residuals; as seen there, larger $\mathcal{B}$, implying the network takes longer time before the union of its instances forms a strongly connected graph, leads to slower convergence. 


\subsubsection{Varied network size}

We now consider the logistic regression problem over networks of varied sizes. In particular, we fix the total number of data points and vary the number of nodes in the network. As the network grows, i.e., the number of nodes in the network becomes larger, each agent has fewer locally available data points. Fig. \ref{fig:expe_more}(b) shows the correct rate for compression levels $q = 1$, $q = 0.12$ and $q = 0.08$ as $n$ grows from $10$ to $30$. For a pre-specified sparsification level, larger networks, in which each agent has fewer data points to train its local model, requires more communication rounds and therefore takes longer to converge. 

\section{Conclusion}\label{sec:conc}

In this paper we studied decentralized convex optimization problems over time-varying directed networks and proposed a stochastic variance-reduced algorithm for solving them. The algorithm sparsifies messages exchanged between network nodes thus enabling collaboration in resource-constrained settings. We proved that the proposed algorithm, Di-CS-SVRG, enjoys linear convergence rate, and demonstrated its efficacy through simulation studies. As part of the future work, it is of interest to extend this work to decentralized non-convex optimization problems.
\clearpage
\printbibliography 
\clearpage
\appendix
\begin{center}
    \textbf{\Large Supplementary Material}\vspace{5mm}
\end{center}
In this document, we include detailed proofs of the auxiliary lemmas stated in Section~1 of the main manuscript, and present experimental results for the consensus algorithm, Algorithm~1.

\section{Analysis}

We start by proving auxiliary lemmas utilized in the proof of Theorem~1. 

\newtheorem*{lemma0}{Lemma $1$}
\begin{lemma0}
Suppose Assumptions~1 (a) and (b) hold. Let $\sigma = \max(|\lambda_{M, 2}|, |\lambda_{B, 2}|)$ denote the larger of the second largest eigenvalues of $M_m((k+1)\mathcal{B}-1:k\mathcal{B} )$ and $B_m((k+1)\mathcal{B}-1:k\mathcal{B})$. Then, 
\begin{align}
\begin{split}
    \|M_m((k+1)\mathcal{B}-1:k\mathcal{B} ) \z - \Bar{\z} \| \leq \sigma \|\z - \Bar{\z} \|, \ \forall \z \in \R^{2n} \\
    \mbox{and}
    \\
    \|B_m((k+1)\mathcal{B}-1:k\mathcal{B} ) \y - \Bar{\y} \| \leq \sigma \|\y - \Bar{\y} \|, \ \forall \y \in \R^{n}, \\
\end{split}
\end{align}
where $\Bar{\z} = [\frac{1}{n}\sum_{i=1}^{2n}z_i, \cdots, \frac{1}{n}\sum_{i=1}^{2n}z_i]^T $ and $\Bar{\y} = [\frac{1}{n}\sum_{i=1}^{n}y_i, \cdots, \frac{1}{n}\sum_{i=1}^{n}y_i]^T $.
\end{lemma0}
\begin{proof}
To prove Lemma~1, we first need to establish the following.

\newtheorem*{lemma01}{Lemma $1.1$}
\begin{lemma01}\label{lemma11}
Assume that $M_m((s+1)\mathcal{B}-1:s\mathcal{B})$ has non-zero spectral gap for each $m$.
Then the following statements hold:
\begin{enumerate}[(a)]
\item The sequence of matrix products $M_m((s+1)\mathcal{B}-1:s\mathcal{B})$ converges to the limit matrix
\begin{equation}
\mathrm{lim}_{t \to \infty} (M_m((s+1)\mathcal{B}-1:s\mathcal{B}))^t=
\left[\begin{matrix}
\frac{\mathbf{1}_n \mathbf{1}^T_n}{n} & \frac{\mathbf{1}_n \mathbf{1}^T_n}{n} \\
\mathbf{0} & \mathbf{0} \\
\end{matrix}
\right].
\end{equation}
\item Let $1=|\lambda_1(M_m((s+1)\mathcal{B}-1:s\mathcal{B}))|>|\lambda_2(M_m((s+1)\mathcal{B}-1:s\mathcal{B}))|\geq \cdots \geq |\lambda_{2n}(M_m((s+1)\mathcal{B}-1:s\mathcal{B}))|$ be the eigenvalues of $M_m((s+1)\mathcal{B}-1:s\mathcal{B})$, and let $\sigma_m=|\lambda_2(M_m((s+1)\mathcal{B}-1:s\mathcal{B}))|$; then there exists $\Gamma_m'>0$ such that
\begin{equation}
\begin{aligned}
\|(M_m((s+1)\mathcal{B}-1:s\mathcal{B}))^t-\mathcal{I}\|_{\infty} 
\leq \Gamma_m' \sigma_m^{t},
\end{aligned}
\end{equation}
where $\mathcal{I}:=\frac{1}{n}[\mathbf{1}^T \ \mathbf{0}^T]^T[\mathbf{1}^T \  \mathbf{1}^T]$.
\end{enumerate}
\end{lemma01}
	
\begin{proof}
For each $m$, $M_m((s+1)\mathcal{B}-1:s\mathcal{B})$ has column sum equal to $1$. According to Assumption~1, definition of the mixing matrix and the construction of the product,
$M_m((s+1)\mathcal{B}-1:s\mathcal{B})$ has a simple eigenvalue $1$ with the corresponding left eigenvector $[\mathbf{1}^T \ \mathbf{1}^T]$ and right eigenvector $[\mathbf{1}^T \ \mathbf{0}^T]^T$.
Following Jordan matrix decomposition for the simple eigenvalue, there exist some $P, Q \in \mathcal{R}^{(2n-1)\times (2n-1)}$ such that
	\begin{equation}
	\begin{aligned}
	(M_m((s+1)\mathcal{B}-1:s\mathcal{B}))^t & =\mathcal{I}^t+P J_m^t Q
	=\mathcal{I}+P J_m^t Q. 
	\end{aligned}
\end{equation}
Let $\gamma_m$ be the second largest eigenvalue magnitude of $M_m((s+1)\mathcal{B}-1:s\mathcal{B})$; then, $\gamma_m$ is also the spectral norm of $J_m$. The proof of part (a) follows by noting that 
$
\mathrm{lim}_{t \to \infty}J_m^t = \mathbf{0}.
$
Since $\|P \|$, $\| Q \|$ and $\|J_m \|$ are finite, there exists some $\Gamma_m' >0$ such that
\begin{equation}
\begin{aligned}
& \quad \|(M_m((s+1)\mathcal{B}-1:s\mathcal{B}))^t-\mathcal{I} \|_{\infty} 
\leq \|P J_m^t Q \|_{\infty} 
\leq \Gamma_m' \sigma_m^t
\end{aligned}
\end{equation}
which completes the proof of part (b).
\end{proof}

Then let $\sigma' = \max_m \sigma_m$, where $\sigma_m$ is as defined in Lemma \ref{lemma11} and, by mathematical induction, for each $m$ it holds that
\begin{equation}
\rho (M_m(T\mathcal{B}-1:0)-\frac{1}{n}[\mathbf{1}^T \ \mathbf{0}^T]^T[\mathbf{1}^T \  \mathbf{1}^T]) \leq  \sigma'^{T}.
\end{equation}

Referring to the fact that $M_m((k+1)\mathcal{B}-1:k\mathcal{B} )$ and $B_m((k+1)\mathcal{B}-1:k\mathcal{B} )$ both have column sums equal to $1$ and defining $\sigma$ as stated in the lemma, we can conclude the proof of Lemma~1.

\end{proof}

The following lemma, restated for convenience, establishes an upper bound on the consensus error.

\newtheorem*{lemma1}{Lemma $2$}
\begin{lemma1}\label{lemma2}
Suppose Assumption~1 holds. Then, $\forall i \leq n$, $k \geq 0$, and $0 < m \leq d$, the updates generated by Algorithm~1 satisfy
 
\begin{equation}
\begin{aligned}
    \mathbb{E}[|z_{im}^{(k+1)\mathcal{B}}-\Bar{z}_{m}^{(k+1)\mathcal{B}} |^2] & \leq \frac{1+\sigma^2}{2}\mathbb{E}[|z_{im}^{k\mathcal{B}}-\Bar{z}_{m}^{k\mathcal{B}} |^2 ] \\ & \quad +\frac{2\alpha^2}{1-\sigma^2}\mathbb{E}[|g_{im}^{k\mathcal{B}}-\Bar{g }_{m}^{k\mathcal{B}} \|^2 ].
\end{aligned}
\end{equation}
\end{lemma1}

\begin{proof}
By constructing normalized weight matrices and relying on the definition of $M_m((k+1)\mathcal{B}-1:k\mathcal{B} ) $, we can simplify the update as
\begin{equation}
\begin{aligned}
    z_{im}^{(k+1)\mathcal{B}} & =\sum_{j=1}^{2n}[M_m((k+1)\mathcal{B}-1:k\mathcal{B} )]_{ij} [Q(z_{j}^{k\mathcal{B}})]_m-\alpha g_{im}^{k\mathcal{B}} \\ & = \sum_{j=1}^{2n}[M_m((k+1)\mathcal{B}-1:k\mathcal{B} )]_{ij} z_{jm}^{k\mathcal{B}}-\alpha g_{im}^{k\mathcal{B}}.
\end{aligned}
\end{equation}
Since for any $m$ we have $\Bar{g }_{m}^{k\mathcal{B}} = \frac{1}{n}\sum_{j=1}^n  g_{jm}^{k\mathcal{B}}$, it holds that
\begin{equation}
\begin{aligned}
    |z_{im}^{(k+1)\mathcal{B}}-\Bar{z}_{m}^{(k+1)\mathcal{B}} |^2 = |\sum_{j=1}^{2n}[M_m((k+1)\mathcal{B}-1:k\mathcal{B} )]_{ij} z_{jm}^{k\mathcal{B}}-\Bar{z}_{m}^{k\mathcal{B}} -\alpha ( g_{im}^{k\mathcal{B}}-\Bar{g }_{m}^{k\mathcal{B}} ) |^2.
\end{aligned}
\end{equation}
By Young's inequality,
\begin{equation*}
    \|\a +\b \|^2 \leq (1+\eta)\|\a \|^2+(1+\frac{1}{\eta})\|\b \|^2,\quad \forall \eta>0, \a, \b.
\end{equation*}
Using Lemma~1, $\forall i \leq n$ and $0 < m \leq d$ it holds that
\begin{equation}
    |\sum_{j=1}^{2n}[M_m((k+1)\mathcal{B}-1:k\mathcal{B} )]_{ij} z_{jm}^{k\mathcal{B}}-\Bar{z}_{m}^{k\mathcal{B}} | \leq \sigma |z_{im}^{k\mathcal{B}}-\Bar{z}_{m}^{k\mathcal{B}} |.
\end{equation}
Then for all $k \geq 0$, 
\begin{equation}
\begin{aligned}
    |z_{im}^{(k+1)\mathcal{B}}-\Bar{z}_{m}^{(k+1)\mathcal{B}} |^2 & \leq (1+\eta)\sigma^2 |z_{im}^{k\mathcal{B}}-\Bar{z}_{m}^{k\mathcal{B}} |^2 + (1+\frac{1}{\eta}) \alpha^2 | g_{im}^{k\mathcal{B}}-\Bar{g }_{m}^{k\mathcal{B}} |^2.
\end{aligned}
\end{equation}
Setting $\eta = \frac{1-\sigma^2}{2\sigma^2}$ completes the proof.

\end{proof}

Next, we provide proofs of two lemmas stating upper bounds on the optimality gap and the gradient tracking error. For convenience, we first introduce 
\begin{equation}
    \Bar{\tau}^{k\mathcal{B}} = \frac{1}{n}\sum_{i=1}^n \tau_i^{k\mathcal{B}}
\end{equation}
 and 
\begin{equation}
    \tau_i^{(k+1)\mathcal{B}} = \begin{cases}
    \x_i^{(k+1)\mathcal{B}} & \mathrm{if} \quad (k+1)\mathcal{B} \mod T = 0 \\
    \tilde{w}_i & \mathrm{otherwise}.
    \end{cases}
\end{equation}


\newtheorem*{lemma3}{Lemma $3$}
\begin{lemma3}
Suppose Assumption~1 holds and
let $0 < \alpha < \frac{\mu}{8L^2}$. Then for all $k>0$ it holds that
\begin{align}
    \begin{split}
    \mathbb{E}[n \|\Bar{\z}^{(k+1)\mathcal{B}} - \x^* \|^2] & \leq  \frac{2L^2 \alpha}{\mu}\mathbb{E}[\sum_{i=1}^n  \|\Bar{\z}^{k\mathcal{B}}-\z_i^{k\mathcal{B}} \|^2 ] + (1-\frac{ \mu \alpha}{2} )\mathbb{E}[n\|\Bar{\z}^{k\mathcal{B}}-\x^* \|^2 ] \\
    & \quad + \frac{4L^2 \alpha^2}{n} \mathbb{E}[ \sum_{i=1}^n \| \tau_i^{k\mathcal{B}}-\Bar{\tau}^{k\mathcal{B}} \|^2] + \frac{4L^2 \alpha^2}{n} \mathbb{E}[n \|\Bar{\tau}^{k\mathcal{B}} - \x^* \|^2 ].
\end{split}
\end{align}
\end{lemma3}

\newtheorem*{lemma4}{Lemma $4$}
\begin{lemma4}
Suppose Assumption~1 holds. Then,
\begin{equation}
\begin{aligned}
\frac{1 }{L^2}\mathbb{E}[\sum_{m=1}^d \sum_{i=1}^n |g_{im}^{(k+1)\mathcal{B}}-\Bar{g}_m^{(k+1)\mathcal{B}} |^2]  & \leq \frac{120}{1-\sigma^2}\mathbb{E}[\sum_{i=1}^n  \|\Bar{\z}^{k\mathcal{B}}-\z_i^{k\mathcal{B}} \|^2  ] + \frac{89}{1-\sigma^2}\mathbb{E}[ n\|\Bar{\z}^{k\mathcal{B}}-\x^* \|^2] \\& \quad + \frac{3+\sigma^2}{4} \mathbb{E}[\frac{\sum_{m=1}^d \sum_{i=1}^n |g_{im}^{k\mathcal{B}}-\Bar{g}_m^{k\mathcal{B}} |^2  }{L^2} ]\\ & \quad + \frac{38}{1-\sigma^2} \mathbb{E}[\sum_{i=1}^n  \|\Bar{\z}^{k\mathcal{B}}-\z_i^{k\mathcal{B}} \|^2  ]
+ \frac{38}{1-\sigma^2} \mathbb{E}[n \|\Bar{\tau}^{k\mathcal{B}} - \x^* \|^2 ].
\end{aligned}
\end{equation}
\end{lemma4}

Proving Lemmas~3 and 4 requires a series of auxiliary lemmas, Lemma~3.1-3.4. We start with Lemma~3.1, which states an upper bound on $\mathbb{E}[\|\Bar{\z}^{(k+1)\mathcal{B}}-\x^* \|^2 ]$.
\newtheorem*{lemma31}{Lemma $3.1$}
\begin{lemma31}
Suppose Assumption~1 holds. 
Let $0 < \alpha < \frac{1}{L}$, where $L$ is the smoothness parameter. For all $k>0$, it holds that
\begin{equation}
\begin{aligned}
    \mathbb{E}[\|\Bar{\z}^{(k+1)\mathcal{B}} - \x^* \|^2] & \leq \frac{L^2\alpha}{n\mu} \mathbb{E}[\sum_{i=1}^n  \|\Bar{\z}^{k\mathcal{B}}-\z_i^{k\mathcal{B}} \|^2 ]  + (1-\mu \alpha)\mathbb{E}[\|\Bar{\z}^{k\mathcal{B}}-\x^* \|^2] + \frac{\alpha^2}{n^2}\mathbb{E}[\|\v^{k\mathcal{B}} -\nabla \mathbf{f}(\x^{k\mathcal{B}}) \|^2  ],
\end{aligned}
\end{equation}
where $\mu$ is the strong convexity parameter and $\nabla \mathbf{f}(\x^{k\mathcal{B}}) = [\nabla f_1(\x_1^{k\mathcal{B}}); \cdots; \nabla f_n(\x_n^{k\mathcal{B}})] $.
\end{lemma31}

\begin{proof}
By definition, $\Bar{z}_{m}^{t} = \frac{1}{n}\sum_{i=1}^{2n} z_{im}^t $. Let us denote $\nabla \Bar{f}(\x^t) = \nabla \Bar{f}(\z^t) = \frac{1}{n}\sum_{i=1}^n \nabla f_i(\z_i^t) $. 
By induction, 
\begin{equation}
    \Bar{g}_m^{k\mathcal{B}} = \Bar{v}_m^{k\mathcal{B}}.
\end{equation}
Next, we have that for any $m$
\begin{equation}
    \Bar{z}_{m}^{(k+1)\mathcal{B}} = \Bar{z}_{m}^{k\mathcal{B}}-\alpha \Bar{g}_m^{k\mathcal{B}} = \Bar{z}_{m}^{k\mathcal{B}}-\alpha \Bar{v}_m^{k\mathcal{B}},
\end{equation}
which implies that
\begin{equation}
    \Bar{\z}^{(k+1)\mathcal{B}} = \Bar{\z}^{k\mathcal{B}}-\alpha \Bar{\g}^{k\mathcal{B}} = \Bar{\z}^{k\mathcal{B}}-\alpha \Bar{\v}^{k\mathcal{B}}.
\end{equation}

Note that the randomness in Algorithm~1 originates from a set of independent random variables $\{\omega_i^t \}_{i \in [2n]}^{t \geq 0}$. We rely on the $\sigma$-algebra $\F^{k\mathcal{B}}$ to characterize the history of the dynamical system generated by $ \{\omega_i^t \}_{i \in [2n]}^{t \leq k\mathcal{B}-1}$,
\begin{align}
\begin{split}
    & \mathbb{E}[\|\Bar{\z}^{(k+1)\mathcal{B}} - \x^* \|^2 | \F^{k\mathcal{B}}] = \mathbb{E}[\|\Bar{\z}^{k\mathcal{B}} -\alpha \Bar{\v}^{k\mathcal{B}}- \x^* \|^2 | \F^{k\mathcal{B}}] \\
    & = \mathbb{E}[\|\Bar{\z}^{k\mathcal{B}}-\alpha \nabla f(\Bar{\z}^{k\mathcal{B}})-\x^* + \alpha (\nabla f(\Bar{\z}^{k\mathcal{B}})- \Bar{\v}^{k\mathcal{B}}) \|^2 | \F^{k\mathcal{B}}] \\
    & = \|\Bar{\z}^{k\mathcal{B}}-\alpha \nabla f(\Bar{\z}^{k\mathcal{B}})-\x^* \|^2 +\alpha^2 \mathbb{E}[\|\nabla f(\Bar{\z}^{k\mathcal{B}})- \Bar{\v}^{k\mathcal{B}}  \|^2 | \F^{k\mathcal{B}} ] \\
    & \quad +2\alpha \langle \Bar{\z}^{k\mathcal{B}}-\alpha \nabla f(\Bar{z}^{k\mathcal{B}})-\x^*, \nabla f(\Bar{\z}^{k\mathcal{B}})- \Bar{\v}^{k\mathcal{B}}  \rangle \\
    & = \|\Bar{\z}^{k\mathcal{B}}-\alpha \nabla f(\Bar{\z}^{k\mathcal{B}})-\x^* \|^2 +\alpha^2 \mathbb{E}[\|\nabla f(\Bar{\z}^{k\mathcal{B}})- \Bar{\v}^{k\mathcal{B}}  \|^2 | \F^{k\mathcal{B}} ] \\
    & \quad +2\alpha \langle \Bar{\z}^{k\mathcal{B}}-\alpha \nabla f(\Bar{\z}^{k\mathcal{B}})-\x^*, \nabla f(\Bar{\z}^{k\mathcal{B}})- \nabla \Bar{f}(\x^{\mathcal{B}})  \rangle.
\end{split}
\end{align}
We then proceed by considering $\mathbb{E}[ \|\nabla f(\Bar{\z}^{k\mathcal{B}})- \Bar{\v}^{k\mathcal{B}}  \|^2 | \F^{k\mathcal{B}} ]$,
\begin{align}
\begin{split}
 \mathbb{E}[ \|\nabla f(\Bar{\z}^{k\mathcal{B}})- \Bar{\v}^{k\mathcal{B}}  \|^2 | \F^{k\mathcal{B}} ] & = \mathbb{E}[ \|\nabla f(\Bar{\z}^{k\mathcal{B}})-\nabla \Bar{f}(\x^{k\mathcal{B}}) +\nabla \Bar{f}(\x^{k\mathcal{B}}) - \Bar{\v}^{k\mathcal{B}}  \|^2 | \F^{k\mathcal{B}} ] \\
    & = \| \nabla f(\Bar{\z}^{k\mathcal{B}})-\nabla \Bar{f}(\x^{k\mathcal{B}})\|^2 + \mathbb{E}[ \|\nabla \Bar{f}(\x^{k\mathcal{B}}) - \Bar{\v}^{k\mathcal{B}}  \|^2|\F^{k\mathcal{B}}],
\end{split}
\end{align}
where the fact that $\mathbb{E}[\Bar{\v}^t |\F^t] = \nabla \Bar{f}(\x^t) $ is used. Furthermore, note that
\begin{equation}
\begin{aligned}
    \mathbb{E}[ \|\nabla \Bar{f}(\x^{k\mathcal{B}}) - \Bar{\v}^{k\mathcal{B}}  \|^2|\F^{k\mathcal{B}}] & = \frac{1}{n^2} \mathbb{E}[\| \sum_{i=1}^n (\v_i^{k\mathcal{B}} - \nabla f_i(\z_i^{k\mathcal{B}})) \|^2 |\F^{k\mathcal{B}} ] \\
    & = \frac{1}{n^2}\mathbb{E}[ \|\v^{k\mathcal{B}} -\nabla \mathbf{f}(\x^{k\mathcal{B}}) \|^2 |\F^{k\mathcal{B}}],
\end{aligned}
\end{equation}
since $\left\{\v_i^t \right\}_{i=1}^n$ are independent given $\F^t$ and $ \mathbb{E}[\sum_{i \neq j}\langle \v_i^t - \nabla f_i(\z_i^t), \v_j^t - \nabla f_j(\z_j^t) \rangle |\F^t]=0$. 
Recall the strong convexity of the objective, i.e., we have that if $0 < \alpha \leq \frac{1}{L}$, $\forall \x$
\begin{equation}
    \| \x - \alpha \nabla \f(\x) -\x^* \| \leq (1-\mu \alpha) \|\x - \x^* \|.
\end{equation}
It follows that
\begin{align}
\begin{split}
    \mathbb{E}[\|\Bar{\z}^{(k+1)\mathcal{B}} - \x^* \|^2 | \F^{k\mathcal{B}}] & \leq (1-\mu \alpha)^2 \|\Bar{z}^{k\mathcal{B}}-\x^* \|^2  + \alpha^2 \| \nabla f(\Bar{\z}^{k\mathcal{B}})-\nabla \Bar{f}(\x^{k\mathcal{B}})\|^2 \\
    & \quad + 2\alpha(1-\mu \alpha)\|\Bar{z}^{k\mathcal{B}}-\x^* \| \| \nabla f(\Bar{\z}^{k\mathcal{B}})-\nabla \Bar{f}(\x^{k\mathcal{B}})\|\\ & \quad + \frac{\alpha^2}{n^2} \mathbb{E}[ \|\v^{k\mathcal{B}} -\nabla \f(\x^{k\mathcal{B}}) \|^2 |\F^{k\mathcal{B}}].
\end{split}
\end{align}
Using Young's inequality, we readily obtain that
\begin{equation}
\begin{aligned}
    2\alpha\|\Bar{\z}^{k\mathcal{B}}-\x^* \|\| \nabla f(\Bar{\z}^{k\mathcal{B}})-\nabla \Bar{f}(\x^{k\mathcal{B}})\| \leq \mu\alpha \|\Bar{\z}^{k\mathcal{B}}-\x^* \|^2  + \frac{\alpha}{\mu}\| \nabla f(\Bar{\z}^{k\mathcal{B}})-\nabla \Bar{f}(\x^{k\mathcal{B}})\|^2.
\end{aligned}
\end{equation}
On the other hand, assuming convexity and smoothness, we have that $\forall k \geq 0$, 
\begin{align}
    \| \nabla f(\Bar{\z}^{k\mathcal{B}}) - \nabla \Bar{f}(\x^{k\mathcal{B}})\| & = 
    \| \sum_{i=1}^n \frac{\nabla f_i(\Bar{\z}^{k\mathcal{B}})-\nabla f_i(\z_i^{k\mathcal{B}})}{n} \| \\
    & \leq L \sum_{i=1}^n \frac{ \|\Bar{\z}^{k\mathcal{B}}-\z_i^{k\mathcal{B}} \|}{n} \\
    & \leq L \sqrt{\sum_{i=1}^n \frac{ \|\Bar{\z}^{k\mathcal{B}}-\z_i^{k\mathcal{B}} \|^2}{n} } \\
    & = \frac{L}{\sqrt{n}} \sqrt{\sum_{i=1}^n  \|\Bar{\z}^{k\mathcal{B}}-\z_i^{k\mathcal{B}} \|^2 }.
\end{align}

Then by taking the total expectation, 
\begin{equation*}
\begin{aligned}
    \mathbb{E}[\|\Bar{\z}^{(k+1)\mathcal{B}} - \x^* \|^2] & \leq \frac{L^2\alpha}{n\mu} \mathbb{E}[\sum_{i=1}^n  \|\Bar{\z}^{k\mathcal{B}}-\z_i^{k\mathcal{B}} \|^2 ] + (1-\mu \alpha)\mathbb{E}[\|\Bar{\z}^{k\mathcal{B}}-\x^* \|^2] + \frac{\alpha^2}{n^2}\mathbb{E}[\|\v^{k\mathcal{B}} -\nabla \mathbf{f}(\x^{k\mathcal{B}}) \|^2  ].
\end{aligned}
\end{equation*}
\end{proof}

The following lemma helps establish an upper bound on the expected gradient tracking error. \newtheorem*{lemma32}{Lemma $3.2$}
\begin{lemma32}\label{lemm3}
Suppose the objective function $f$ is $\mu$-strongly-convex and that each component of the local objective function $f_{i, j}$ is $L$-smooth. If $0 <\alpha < \frac{1}{4\sqrt{2}L}$, 
\begin{align}
\begin{split}
    \mathbb{E}[\|\g_{\sim n}^{(k+1)\mathcal{B}} 
     - \mathbf{1}_{n} \Bar{\g}^{(k+1)\mathcal{B}} \|^2 ] & \leq \frac{33L^2}{1-\sigma^2} E[\|\z^{k\mathcal{B}}-\mathbf{1}_{2n}(\Bar{\z}^{k\mathcal{B}})' \|^2] + \frac{2L^2}{1-\sigma^2}\mathbb{E}[n\|\Bar{\z}^{k\mathcal{B}} - \x^* \|^2 ] \\ & \quad +(\frac{1+\sigma^2}{2}+\frac{32  \alpha^2L^2}{1-\sigma^2}) \mathbb{E}[\|\g_{\sim n}^{k\mathcal{B}} 
     - \mathbf{1}_{n} \Bar{\g}^{k\mathcal{B}} \|^2 ]\\
     & \quad + \frac{5}{1-\sigma^2} \mathbb{E}[\|\v^{k\mathcal{B}} - \nabla \f (\x^{k\mathcal{B}}) \|^2 ] \\ & \quad + \frac{4}{1-\sigma^2} \mathbb{E}[\|\v^{(k+1)\mathcal{B}}-\nabla \f(\x^{(k+1)\mathcal{B}}) \|^2 ],
\end{split}
\end{align}
where $\g_{\sim n}^t = [\g_1^t; \cdots; \g_n^t] \in \R^{n \times d}.$
\end{lemma32}

\begin{proof}
For all $i \leq n$ and $0 < m \leq d$, it holds that
\begin{equation}
\begin{aligned}
    |g_{im}^{(k+1)\mathcal{B}}-\Bar{g}_m^{(k+1)\mathcal{B}} |^2 & = |\sum_{j=1}^{n}[B_m((k+1)\mathcal{B}-1:k\mathcal{B} )]_{ij} g_{jm}^{k\mathcal{B}} + v_{im}^{(k+1)\mathcal{B}}-v_{im}^{k\mathcal{B}}\\ & \quad - \frac{1}{n}\sum_{l=1}^n (\sum_{j=1}^{n}[B_m((k+1)\mathcal{B}-1:k\mathcal{B} )]_{lj} g_{jm}^{k\mathcal{B}} + v_{lm}^{(k+1)\mathcal{B}}-v_{lm}^{k\mathcal{B}} ) |^2.
\end{aligned}
\end{equation}
Denoting $\g_{:m}^{(k+1)\mathcal{B}} = [g_{1m}^{(k+1)\mathcal{B}}, \cdots, g_{nm}^{(k+1)\mathcal{B}} ]^T $ and $\v_{:m}^{(k+1)\mathcal{B}} = [v_{1m}^{(k+1)\mathcal{B}}, \cdots, v_{nm}^{(k+1)\mathcal{B}}] $,
\begin{align}
\begin{split}
    \sum_{i=1}^n |g_{im}^{(k+1)\mathcal{B}}-\Bar{g}_m^{(k+1)\mathcal{B}} |^2 & = \|\g_{:m}^{(k+1)\mathcal{B}}- \Bar{g}_m^{(k+1)\mathcal{B}} \mathbf{1}_n \|^2 \\
    & = \sum_{i=1}^n |\sum_{j=1}^{n}[B_m((k+1)\mathcal{B}-1:k\mathcal{B} )]_{ij} g_{jm}^{k\mathcal{B}} + v_{im}^{(k+1)\mathcal{B}}-v_{im}^{k\mathcal{B}} - \\& \quad \frac{1}{n}\sum_{l=1}^n (\sum_{j=1}^{n}[B_m((k+1)\mathcal{B}-1:k\mathcal{B} )]_{lj} g_{jm}^{k\mathcal{B}} + v_{lm}^{(k+1)\mathcal{B}}-v_{lm}^{k\mathcal{B}} ) |^2 \\
    & = \| B_m((k+1)\mathcal{B}-1:k\mathcal{B} ) \g_{:m}^{k\mathcal{B}} - \Bar{g}_m^{k\mathcal{B}} \mathbf{1}_n  +  (\v_{:m}^{(k+1)\mathcal{B}}-\Bar{v}_m^{(k+1)\mathcal{B}}\mathbf{1}_n ) -(\v_{:m}^{k\mathcal{B}}-\Bar{v}_m^{k\mathcal{B}}\mathbf{1}_n  ) \|^2.
\end{split}
\end{align}
Once again applying Young's inequality yields
\begin{equation}
\begin{aligned}
    \sum_{i=1}^n |g_{im}^{(k+1)\mathcal{B}}-\Bar{g}_m^{(k+1)\mathcal{B}} |^2  & \leq (1+\frac{1-\sigma^2}{2\sigma^2})\| B_m((k+1)\mathcal{B}-1:k\mathcal{B} ) \g_{:m}^{k\mathcal{B}} - \Bar{g}_m^{k\mathcal{B}} \mathbf{1}_n \|^2\\ &\quad +(1+\frac{2\sigma^2}{1-\sigma^2}) \| (\v_{:m}^{(k+1)\mathcal{B}}-\Bar{v}_m^{(k+1)\mathcal{B}}\mathbf{1}_n )  -(\v_{:m}^{k\mathcal{B}}-\Bar{v}_m^{k\mathcal{B}}\mathbf{1}_n  ) \|^2.
\end{aligned}
\end{equation}
Summing up the above objects over all $m$, we obtain
\begin{align}
\begin{split}
    \sum_{m=1}^d \sum_{i=1}^n |g_{im}^{(k+1)\mathcal{B}}-\Bar{g}_m^{(k+1)\mathcal{B}} |^2  & \leq \sum_{m=1}^d (1+\frac{1-\sigma^2}{2\sigma^2})\| B_m((k+1)\mathcal{B}-1:k\mathcal{B} ) \g_{:m}^{k\mathcal{B}} - \Bar{g}_m^{k\mathcal{B}} \mathbf{1}_n \|^2\\ & \quad +(1+\frac{2\sigma^2}{1-\sigma^2}) \| (\v_{:m}^{(k+1)\mathcal{B}}-\Bar{v}_m^{(k+1)\mathcal{B}}\mathbf{1}_n )-(\v_{:m}^{k\mathcal{B}}-\Bar{v}_m^{k\mathcal{B}}\mathbf{1}_n  ) \|^2 \\ 
    & \leq \sum_{m=1}^d \frac{1+\sigma^2}{2} \|\g_{:m}^{k\mathcal{B}} - \Bar{g}_m^{t} \mathbf{1}_n \|^2 \\ & \quad + \frac{2}{1-\sigma^2} \| (\v_{:m}^{(k+1)\mathcal{B}}-\Bar{v}_m^{(k+1)\mathcal{B}}\mathbf{1}_n )-(\v_{:m}^{k\mathcal{B}}-\Bar{v}_m^{k\mathcal{B}}\mathbf{1}_n  ) \|^2 \\
    & \leq  \frac{1+\sigma^2}{2} \sum_{m=1}^d \|\g_{:m}^{k\mathcal{B}} - \Bar{g}_m^{k\mathcal{B}} \mathbf{1}_n \|^2 + \frac{2}{1-\sigma^2}\| \v^{(k+1)\mathcal{B}}-\v^{k\mathcal{B}} \|^2
\end{split}
\end{align}
Taking the total expectation yields
\begin{equation}\label{lemma32_last}
\begin{aligned}
    \mathbb{E}[\sum_{m=1}^d \sum_{i=1}^n |g_{im}^{(k+1)\mathcal{B}}-\Bar{g}_m^{(k+1)\mathcal{B}} |^2 ]& \leq \frac{1+\sigma^2}{2}\mathbb{E}[\sum_{m=1}^d \|\g_{:m}^{k\mathcal{B}} - \Bar{g}_m^{k\mathcal{B}} \mathbf{1}_n \|^2 ] +\frac{2}{1-\sigma^2}\mathbb{E}[\| \v^{(k+1)\mathcal{B}}-\v^{k\mathcal{B}} \|^2 ].
\end{aligned}
\end{equation}

Next, we derive an upper bound on $\mathbb{E}[\| \v^{(k+1)\mathcal{B}}-\v^{k\mathcal{B}} \|^2  ]$ as
\begin{align}
\begin{split}
    \mathbb{E}[\| \v^{(k+1)\mathcal{B}}-\v^{k\mathcal{B}} \|^2 ] & \leq 2\mathbb{E}[\| \v^{(k+1)\mathcal{B}}-\v^{k\mathcal{B}} - (\nabla \f(\x^{(k+1)\mathcal{B}}) - \nabla \f(\x^{k\mathcal{B}})) \|^2 ] \\ & \quad + 2 \mathbb{E}[\|\nabla \f(\x^{(k+1)\mathcal{B}}) - \nabla \f(\x^{k\mathcal{B}}) \|^2 ] \\
    & \leq 2\mathbb{E}[\|\v^{(k+1)\mathcal{B}} -\nabla \f(\x^{(k+1)\mathcal{B}}) \|^2 ] \\ & \quad + 2\mathbb{E}[\|\v^{k\mathcal{B}}-\nabla \f(\x^{k\mathcal{B}}) \|^2 ] + 2L^2 \mathbb{E}[\|\x^{(k+1)\mathcal{B}}-\x^{k\mathcal{B}} \|^2] \\
    & \leq 2\mathbb{E}[\|\v^{(k+1)\mathcal{B}}-\nabla \f(\x^{(k+1)\mathcal{B}}) \|^2 ] \\ & \quad + 2\mathbb{E}[\|\v^{k\mathcal{B}}-\nabla \f(\x^{k\mathcal{B}}) \|^2 ] + 2L^2 \mathbb{E}[\|\z^{(k+1)\mathcal{B}}-\z^{k\mathcal{B}} \|^2],
\end{split}
\end{align}
where $\nabla \f(\x^{(k+1)\mathcal{B}}) = [\nabla f_1(\x_1^{(k+1)\mathcal{B}}); \cdots; \nabla f_n(\x_n^{(k+1)\mathcal{B}}) ] $. To proceed, let us derive an upper bound on $E[\|\z^{(k+1)\mathcal{B}}-\z^{k\mathcal{B}} \|^2 ] $. First, consider each column of $\z^{(k+1)\mathcal{B}}$ and $\z^{k\mathcal{B}}$ (i.e., $\z_{:m}^{(k+1)\mathcal{B}}$ and $\z_{:m}^{k\mathcal{B}}$) separately and observe that
\begin{align}
\begin{split}
    \|\z_{:m}^{(k+1)\mathcal{B}}-\z_{:m}^{k\mathcal{B}} \|^2 & = \|M_m((k+1)\mathcal{B}-1:k\mathcal{B} ) \z_{:m}^{k\mathcal{B}} - \alpha \g_{:m}^{k\mathcal{B}} -\z_{:m}^{k\mathcal{B}} \|^2  \leq 8 \|\z_{:m}^{k\mathcal{B}}-\Bar{z}_{m}^{k\mathcal{B}} \mathbf{1}_{2n} \|^2 + 2\alpha^2 \|\g_{:m}^{k\mathcal{B}} \|^2.
\end{split}    
\end{align}
Then
\begin{equation}
    \|\z^{(k+1)\mathcal{B}} -\z^{k\mathcal{B}}\|^2 \leq 8\sum_{m=1}^d \|\z_{:m}^{k\mathcal{B}}-\Bar{z}_{m}^{k\mathcal{B}} \mathbf{1}_{2n} \|^2 + 2\alpha^2 \|\g^{k\mathcal{B}} \|^2.
\end{equation}
To derive an upper bound on $\|\g_{\sim n}^{k\mathcal{B}} \|$, let 
$\Bar{\z}^{k\mathcal{B}}=[\Bar{z}_{1}^{k\mathcal{B}} \mathbf{1}_{2n}, \cdots, \Bar{z}_{d}^{k\mathcal{B}} \mathbf{1}_{2n}] $ and note that
\begin{align}
\begin{split}
    \|\g_{\sim n}^{k\mathcal{B}} \| & = \|\g_{\sim n}^{k\mathcal{B}} - \mathbf{1}_{n} (\Bar{\g}^{k\mathcal{B}})' + \mathbf{1}_{n} (\Bar{\v}^{k\mathcal{B}})' - \mathbf{1}_{n} (\nabla \Bar{\f} (\x^{k\mathcal{B}}))'
    + \mathbf{1}_{n} (\nabla \Bar{\f} (\x^{k\mathcal{B}}))' - \mathbf{1}_{n} (\nabla \Bar{\f} (\x^*))' \| \\
    & \leq \|\g_{\sim n}^{k\mathcal{B}} - \mathbf{1}_{n} (\Bar{\g}^{k\mathcal{B}})' \| + \|\v^{k\mathcal{B}} - \nabla \f (\x^{k\mathcal{B}}) \| + L\|\x^{k\mathcal{B}} - \mathbf{1}_n (\x^*)' \| \\
    & \leq \|\g_{\sim n}^{k\mathcal{B}} - \mathbf{1}_{n} (\Bar{\g}^{k\mathcal{B}})' \| + \|\v^{k\mathcal{B}} - \nabla \f (\x^{k\mathcal{B}}) \| + L\|\x^{k\mathcal{B}} - \mathbf{1}_{n}(\Bar{\z}^{k\mathcal{B}})' + \mathbf{1}_{n}(\Bar{\z}^{k\mathcal{B}})' - \mathbf{1}_{n}(\x^*)' \| \\
    & \leq \|\g_{\sim n}^{k\mathcal{B}} - \mathbf{1}_{n} (\Bar{\g}^{k\mathcal{B}})' \| + \|\v^{k\mathcal{B}} - \nabla \f (\x^{k\mathcal{B}}) \| + L\|\z^{k\mathcal{B}}-\mathbf{1}_{2n}(\Bar{\z}^{k\mathcal{B}})' \|  + \sqrt{2n}L\|\Bar{\z}^{k\mathcal{B}} - \x^* \|.
\end{split}
\end{align}
Squaring both sides of the above inequality yields
\begin{equation}
\begin{aligned}
    \|\g_{\sim n}^{k\mathcal{B}} \|^2& \leq 4L^2\|\z^{k\mathcal{B}}-\mathbf{1}_{2n}(\Bar{\z}^{k\mathcal{B}})' \|^2 + 8nL^2 \|\Bar{\z}^{k\mathcal{B}} - \x^* \|^2 + 4 \|\g_{\sim n}^{k\mathcal{B}} - \mathbf{1}_{n} (\Bar{\g}^{k\mathcal{B}})' \|^2 + 4 \|\v^{k\mathcal{B}} - \nabla \f (\x^{k\mathcal{B}}) \|^2.
\end{aligned}
\end{equation}
Imposing $0 <\alpha < \frac{1}{4\sqrt{2}L}$, 
\begin{equation}
\begin{aligned}
    \mathbb{E}[\|\z^{(k+1)\mathcal{B}} -\z^{k\mathcal{B}}\|^2] & \leq 8.25\mathbb{E}[\|\z^{k\mathcal{B}}-\mathbf{1}_{2n}(\Bar{\z}^{k\mathcal{B}})' \|^2] + 0.5\mathbb{E}[n\|\Bar{\z}^{k\mathcal{B}} - \x^* \|^2 ] \\ & \quad + 8\alpha^2 \mathbb{E}[\|\g_{\sim n}^{k\mathcal{B}} - \mathbf{1}_{n} (\Bar{\g}^{k\mathcal{B}})' \|^2 ] + 8\alpha^2 \mathbb{E}[\|\v^{k\mathcal{B}} - \nabla \f (\x^{k\mathcal{B}}) \|^2 ].
\end{aligned}
\end{equation}
Then 
\begin{align}\label{equa_43}
\begin{split}
    \mathbb{E}[\| \v^{(k+1)\mathcal{B}}-\v^{k\mathcal{B}} \|^2 ]  & \leq 16.5L^2\mathbb{E}[\|\z^{k\mathcal{B}}-\mathbf{1}_{2n}(\Bar{\z}^{k\mathcal{B}})' \|^2] + L^2\mathbb{E}[n\|\Bar{\z}^{k\mathcal{B}} - \x^* \|^2 ] \\& \quad + 16 \alpha^2L^2 \mathbb{E}[\|\g_{\sim n}^{k\mathcal{B}} 
     - \mathbf{1}_{n} (\Bar{\g}^{k\mathcal{B}})' \|^2 ]  + 2.5 \mathbb{E}[\|\v^{k\mathcal{B}} - \nabla \f (\x^{k\mathcal{B}}) \|^2 ] \\ & \quad + 2 \mathbb{E}[\|\v^{(k+1)\mathcal{B}}-\nabla \f(\x^{(k+1)\mathcal{B}}) \|^2 ].
\end{split}
\end{align}

The proof is completed by combining \eqref{lemma32_last} with \eqref{equa_43}.

\end{proof}

The gradient estimate error, $\mathbb{E}[\|\v^{k\mathcal{B}}-\nabla f(\x^{k\mathcal{B}}) \|^2]$, appearing on the right-hand side of the inequalities in Lemmas~3.1 and 3.2, is analyzed in the following lemma.

\newtheorem*{lemma33}{Lemma $3.3$}
\begin{lemma33}\label{lemm33}
Suppose the objective function $f$ is $\mu$-strongly-convex, and let
$\tau$, $\Bar{\tau}$ be defined as above. Then $\forall k \geq 0$, \begin{equation}
\begin{aligned}
    \mathbb{E}[\|\v^{k\mathcal{B}}-\nabla f(\x^{k\mathcal{B}}) \|^2] &  \leq  4L^2\sum_{i=1}^n \mathbb{E}[ \|\x_i^{k\mathcal{B}} -\Bar{\z}^{k\mathcal{B}} \|^2] + 4L^2 \mathbb{E}[n \|\Bar{\z}^{k\mathcal{B}}-\x^* \|^2 ] \\ & \quad + 4L^2\sum_{i=1}^n\mathbb{E}[ \|\tau_i^{k\mathcal{B}} -\Bar{\tau}^{k\mathcal{B}} \|^2] + 4L^2 \mathbb{E}[n\|\Bar{\tau}^{k\mathcal{B}} - \x^* \|^2].
\end{aligned}
\end{equation}
\end{lemma33}
\begin{proof}
For all $i \leq n$, it holds that
\begin{align}
\begin{split}
    \mathbb{E}[\|\v_i^{k\mathcal{B}}-\nabla f_i(\x_i^{k\mathcal{B}}) \|^2|\F^{k\mathcal{B}}] & = \mathbb{E}[\|\nabla f_{i, l_i^{k\mathcal{B}}}(\x_i^{k\mathcal{B}}) - f_{i, l_i^{k\mathcal{B}}}(\tau_i^{k\mathcal{B}}) - (\nabla f_i(\x_i^{k\mathcal{B}}) - \nabla f_i(\tau_i^{k\mathcal{B}}) )\|^2 |\F^{k\mathcal{B}}] \\
    & \leq \mathbb{E}[\|\nabla f_{i, l_i^{k\mathcal{B}}}(\x_i^{k\mathcal{B}}) - f_{i, l_i^{k\mathcal{B}}}(\tau_i^{k\mathcal{B}}) \|^2 |\F^{k\mathcal{B}}] \\
    & = \frac{1}{m_i} \sum_{j=1}^{m_i} \|\nabla f_{i,j}(\x_i^{k\mathcal{B}})-\nabla f_{i, j}(\x^*) + (\nabla f_{i, j}(\x^*) -\nabla f_{i, j}(\tau_i^{k\mathcal{B}}) ) \|^2 \\
    & \leq 2L^2\|\x_i^{k\mathcal{B}} -\x^* \|^2 +2L^2\|\tau_i^{k\mathcal{B}} - \x^* \|^2 \\
    & \leq 4L^2 \|\x_i^{k\mathcal{B}} -\Bar{\z}^{k\mathcal{B}} \|^2 + 4L^2 \|\Bar{\z}^{k\mathcal{B}}-\x^* \|^2 + 4L^2 \|\tau_i^{k\mathcal{B}} -\Bar{\tau}^{k\mathcal{B}} \|^2+ 4L^2 \|\Bar{\tau}^{k\mathcal{B}} - \x^* \|^2.
\end{split}
\end{align}
The proof of the lemma is completed by summing over $i$ from $1$ to $n$ and taking the total expectation. 

\end{proof}

Combining the results of Lemma~2, 3.1 and 3.3, we obtain the following result.
\newtheorem*{lemma34}{Lemma $3.4$}
\begin{lemma34}\label{lemma34}
Suppose the objective function $f$ is $\mu$-strongly-convex. 
If $0 < \alpha \leq \frac{1}{8L}$, then for all $k \geq 0$ it holds
\begin{align}
\begin{split}
    \mathbb{E}[\|\v_i^{(k+1)\mathcal{B}}-\nabla f_i(\x_i^{(k+1)\mathcal{B}}) \|^2] & \leq 16.75L^2 \mathbb{E}[\|\x_i^{(k+1)\mathcal{B}} -\Bar{\z}^{(k+1)\mathcal{B}}  \|^2]\\ & \quad + 16L^2 \alpha^2\mathbb{E}[\|\g_i^{k\mathcal{B}} - \mathbf{1}_n \Bar{g}^{k\mathcal{B}} \|^2] + 16.5L^2 \mathbb{E}[\|\Bar{\z}^{k\mathcal{B}}-\x^* \|^2] \\
    & \quad + 4.5L^2 \mathbb{E}[\|\tau_i^{k\mathcal{B}} -\Bar{\tau}^{k\mathcal{B}} \|^2] + 4.5L^2 \mathbb{E}[\|\Bar{\tau}^{k\mathcal{B}} - \x^* \|^2].
\end{split}
\end{align}
\end{lemma34}

\begin{proof}
The proof is completed by combining Lemma~1, 2 and 3.3.
\end{proof}

We can now present an argument proving Lemmas~3 and 4 in the main paper. In particular, combining Lemma~3.1
\begin{equation*}
\begin{aligned}
    \mathbb{E}[\|\Bar{\z}^{(t+1)\mathcal{B}} - \x^* \|^2] \leq \frac{L^2\alpha}{n\mu} \mathbb{E}[\sum_{i=1}^n  \|\Bar{\z}^{k\mathcal{B}}-\z_i^{k\mathcal{B}} \|^2 ] + (1-\mu \alpha)\mathbb{E}[\|\Bar{z}^{k\mathcal{B}}-\x^* \|^2] + \frac{\alpha^2}{n^2}\mathbb{E}[\|\v^{k\mathcal{B}} -\nabla f(\x^{k\mathcal{B}}) \|^2  ]
\end{aligned}
\end{equation*}
and the results in Lemma~3.2, we obtain
\begin{align}
\begin{split}
    \mathbb{E}[n \|\Bar{\z}^{(k+1)\mathcal{B}} - \x^* \|^2] & \leq L^2 \alpha (\frac{1}{\mu}+ \frac{4\alpha}{n})\mathbb{E}[\sum_{i=1}^n  \|\Bar{\z}^{k\mathcal{B}}-\z_i^{k\mathcal{B}} \|^2 ] + (1-\mu \alpha +\frac{4L^2\alpha^2}{n})\mathbb{E}[n\|\Bar{\z}^{k\mathcal{B}}-\x^* \|^2 ] \\
    & \quad + \frac{4L^2 \alpha^2}{n} \mathbb{E}[ \sum_{i=1}^n \| \tau_i^{k\mathcal{B}}-\Bar{\tau}^{k\mathcal{B}} \|^2] + \frac{4L^2 \alpha^2}{n} \mathbb{E}[n \|\Bar{\tau}^{k\mathcal{B}} - \x^* \|^2 ].
\end{split}
\end{align}
By letting $0 < \alpha \leq \frac{\mu}{8L^2}$, we can further bound
\begin{align*}
    \begin{split}
    \mathbb{E}[n \|\Bar{\z}^{(k+1)\mathcal{B}} - \x^* \|^2] & \leq  \frac{2L^2 \alpha}{\mu}\mathbb{E}[\sum_{i=1}^n  \|\Bar{\z}^{k\mathcal{B}}-\z_i^{k\mathcal{B}} \|^2 ] + (1-\frac{ \mu \alpha}{2} )\mathbb{E}[n\|\Bar{\z}^{k\mathcal{B}}-\x^* \|^2 ] \\
    & \quad + \frac{4L^2 \alpha^2}{n} \mathbb{E}[ \sum_{i=1}^n \| \tau_i^{k\mathcal{B}}-\Bar{\tau}^{k\mathcal{B}} \|^2] + \frac{4L^2 \alpha^2}{n} \mathbb{E}[n \|\Bar{\tau}^{k\mathcal{B}} - \x^* \|^2 ],
\end{split}
\end{align*}
which completes the proof of Lemma~3. Moreover,
\begin{align}
\begin{split}
    \mathbb{E}[\sum_{m=1}^d \sum_{i=1}^n |g_{im}^{(k+1)\mathcal{B}}-\Bar{g}_m^{(k+1)\mathcal{B}} |^2] & \leq \frac{120L^2}{1-\sigma^2}\mathbb{E}[\sum_{i=1}^n  \|\Bar{\z}^{k\mathcal{B}}-\z_i^{k\mathcal{B}} \|^2 ]\\ & \quad + \frac{89L^2}{1-\sigma^2} \mathbb{E}[n\|\Bar{\z}^{k\mathcal{B}}-\x^* \|^2 ] \\
    & \quad + (\frac{1+\sigma^2}{2}+\frac{96L^2\alpha^2}{1-\sigma^2}) \mathbb{E}[\sum_{m=1}^d \sum_{i=1}^n |g_{im}^{k\mathcal{B}}-\Bar{g}_m^{k\mathcal{B}} |^2 ] \\
    & \quad + \frac{38L^2}{1-\sigma^2} \mathbb{E}[\sum_{i=1}^n \| \tau_i^{k\mathcal{B}}-\Bar{\tau}^{k\mathcal{B}} \|^2 ]\\ & \quad + \frac{38L^2}{1-\sigma^2} \mathbb{E}[n \|\Bar{\tau}^{k\mathcal{B}} - \x^* \|^2 ].
\end{split}
\end{align}
For $0 < \alpha \leq \frac{1-\sigma^2}{14\sqrt{2}L} $, we have $ \frac{1+\sigma^2}{2}+\frac{98L^2\alpha^2}{1-\sigma^2} \leq \frac{3+\sigma^2}{4}$; this helps complete the proof of Lemma~4,
\begin{align*}
\begin{split}
    \mathbb{E}[\sum_{m=1}^d \sum_{i=1}^n |g_{im}^{(k+1)\mathcal{B}}-\Bar{g}_m^{(k+1)\mathcal{B}} |^2] & \leq \frac{120L^2}{1-\sigma^2}\mathbb{E}[\sum_{i=1}^n  \|\Bar{\z}^{k\mathcal{B}}-\z_i^{k\mathcal{B}} \|^2 ]\\ & \quad + \frac{89L^2}{1-\sigma^2} \mathbb{E}[n\|\Bar{\z}^{k\mathcal{B}}-\x^* \|^2 ] \\
    & \quad + \frac{3+\sigma^2}{4}\mathbb{E}[\sum_{m=1}^d \sum_{i=1}^n \|g_{im}^{k\mathcal{B}}-\Bar{g}_m^{k\mathcal{B}} \|^2 ] \\
    & \quad + \frac{38L^2}{1-\sigma^2} \mathbb{E}[\sum_{i=1}^n \| \tau_i^{k\mathcal{B}}-\Bar{\tau}^{k\mathcal{B}} \|^2 ]\\ & \quad + \frac{38L^2}{1-\sigma^2} \mathbb{E}[n \|\Bar{\tau}^{k\mathcal{B}} - \x^* \|^2 ].
\end{split}
\end{align*}

Using the inequalities shown in Lemma~2, 3 and 4, we can construct a dynamic system and continue the proof of linear convergence of Algorithm~1. To this end, we first define
\begin{equation}
    \u^{k\mathcal{B}} = \begin{bmatrix}
    \mathbb{E}[\sum_{i=1}^n  \|\Bar{\z}^{k\mathcal{B}}-\z_i^{k\mathcal{B}} \|^2  ] \\
    \mathbb{E}[ n\|\Bar{\z}^{k\mathcal{B}}-\x^* \|^2] \\
    \mathbb{E}[\frac{\sum_{m=1}^d \sum_{i=1}^n \|g_{im}^{k\mathcal{B}}-\Bar{g}_m^{k\mathcal{B}} \|^2  }{L^2} ]
    \end{bmatrix}
\end{equation}

\begin{equation}
    \Tilde{\u}^{k\mathcal{B}} = \begin{bmatrix}
     \mathbb{E}[\sum_{i=1}^n \| \tau_i^{k\mathcal{B}}-\Bar{\tau}^{k\mathcal{B}} \|^2 ] \\
     \mathbb{E}[n \|\Bar{\tau}^{k\mathcal{B}} - \x^* \|^2 ] \\
     \mathbf{0}
    \end{bmatrix}
\end{equation}

\begin{equation}
    J_{\alpha} = \begin{bmatrix}
    \frac{1+\sigma^2}{2} & 0 & \frac{2\alpha^2L^2}{1-\sigma^2} \\
    \frac{2L^2\alpha}{\mu} & 1-\frac{\mu \alpha}{2} & 0 \\
    \frac{120}{1-\sigma^2} & \frac{89}{1-\sigma^2} & \frac{3+\sigma^2}{4} 
    \end{bmatrix}
\end{equation}

\begin{equation}
    H_{\alpha} = \begin{bmatrix}
    0 & 0 & 0 \\
    \frac{4L^2 \alpha^2}{n} & \frac{4L^2 \alpha^2}{n} & 0 \\
    \frac{38}{1-\sigma^2} & \frac{38}{1-\sigma^2} & 0
    \end{bmatrix}
\end{equation}
and then formally state the dynamic system in Proposition~1. 
\begin{proposition}\label{proposition1}
Suppose Assumption~1 holds, the objective function $f$ is $\mu$-strongly-convex and each component of the local objective function $f_{i, j}$ is $L$-smooth. If $0 \leq \alpha \leq \frac{\mu(1-\sigma^2)}{14\sqrt{2}L^2}$, then for any
$k \geq 0$
\begin{equation}
    \u^{(k+1)\mathcal{B}} \leq J_{\alpha}\u^{k\mathcal{B}} + H_{\alpha} \Tilde{\u}^{k\mathcal{B}}.
\end{equation}
\end{proposition}
It follows that for the inner loop, for all $k \in [sT, (s+1)T-1]$
\begin{equation}
    \u^{(k+1)\mathcal{B}} \leq J_{\alpha}\u^{k\mathcal{B}} + H_{\alpha} \u^{sT}.
\end{equation}
For the outer loop, for all $s \geq 0$, it holds
\begin{equation}
    \u^{(s+1)T} \leq (J_{\alpha}^T + \sum_{l=0}^{T-1}J_{\alpha}^l H_{\alpha}) \u^{sT}.
\end{equation}

To guarantee linear decay of the outer loop sequence and ultimately show linear convergence of Algorithm~1, we require that the spectral radius of $J_{\alpha}^T + \sum_{l=0}^{T-1}J_{\alpha}^l H_{\alpha}$ is small. In Lemma~5, we compute the range of the step size, $\alpha$, such that the weighted matrix norms of both $ J_{\alpha}^T $ and $ \sum_{l=0}^{T-1}J_{\alpha}^l H_{\alpha}$ are small.
\newtheorem*{lemma5}{Lemma $5$}
\begin{lemma5}\label{lemma5}

Suppose Assumption~1 holds and assume that $ 0 < \alpha \leq \frac{(1-\sigma^2)^2}{187\Tilde{Q}L}$, where $\Tilde{Q} = \frac{L}{\mu}$.
Then,
\begin{equation}
    \rho(J_{\alpha}) < \| |J_{\alpha}| \|^{\mathbf{\delta}}_{\infty} < 1 - \frac{\mu \alpha}{4},
\end{equation}
and
\begin{equation}
    \| |\sum_{l=0}^{T-1}J_{\alpha}^l H_{\alpha}| \|^{\q}_{\infty} \leq
     \| | (I - J_{\alpha})^{-1} H_{\alpha} | \|^{\q}_{\infty}< 0.66,
\end{equation}
where $\mathbf{\delta} = \begin{bmatrix}
1, 8\Tilde{Q}^2, \frac{6656\Tilde{Q}^2}{(1-\sigma^2)^2}
\end{bmatrix}$ and $\q = [1, 1, \frac{1457}{(1-\sigma^2)^2}]$.

\end{lemma5}

\begin{proof}
Following Lemma~10 in \cite{xin2019distributed}, consider a matrix $A \in \R^{d \times d}$ and a positive vector $\x \in \R^d$, and note that if $A\x \leq \beta \x$ for $\beta >0$ we have that $\rho(A) \leq \| |A| \|^{\x}_{\infty} \leq \beta$. Using this lemma we solve for a range of $\alpha$ and a positive vector $\mathbf{\delta} \in \R^3$ such that
\begin{equation}
    J_{\alpha} \mathbf{\delta} \leq (1-\frac{\mu \alpha}{4}) \mathbf{\delta},
\end{equation}
which is equivalent to the element-wise inequalities
\begin{equation}
\begin{aligned}
    \frac{1+\sigma^2}{2}\delta_1 +  \frac{2\alpha^2L^2}{1-\sigma^2}\delta_3 \leq (1-\frac{\mu \alpha}{4}) \delta_1 \\
    \frac{2L^2\alpha}{\mu}\delta_1 + (1-\frac{\mu \alpha}{2})\delta_2 \leq (1-\frac{\mu \alpha}{4}) \delta_2 \\
    \frac{120}{1-\sigma^2}\delta_1 + \frac{89}{1-\sigma^2}\delta_2 + \frac{3+\sigma^2}{4} \delta_3 \leq (1-\frac{\mu \alpha}{4})\delta_3.
\end{aligned}
\end{equation}
To solve for a meaningful $\mathbf{\delta}$, we set $\delta_1 = 1$ and $\delta_2 = 8\Tilde{Q}^2$; then $\mathbf{\delta} = \begin{bmatrix}
1, 8\Tilde{Q}^2, \frac{6656\Tilde{Q}^2}{(1-\sigma^2)^2}
\end{bmatrix} $ and $0 < \alpha \leq \frac{(1-\sigma^2)^2}{187\Tilde{Q}L} $ are sufficient to satisfy the first inequality.
Since $J_{\alpha}$ is non-negative,
$\sum_{l=0}^{T-1}J_{\alpha}^l \leq \sum_{l=0}^{\infty}J_{\alpha}^l = (I_3 - J_{\alpha})^{-1}$; this yields
\begin{equation}
    \u^{(s+1)T} \leq (J_{\alpha}^T + (I_3 - J_{\alpha})^{-1} H_{\alpha}) \u^{sT},
\end{equation}
where
\begin{equation}
    I_3 - J_{\alpha} = \begin{bmatrix}
    \frac{1-\sigma^2}{2} & 0 & -\frac{2\alpha^2L^2}{1-\sigma^2} \\
    -\frac{2L^2\alpha}{\mu} & \frac{\mu \alpha}{2} & 0 \\
    -\frac{120}{1-\sigma^2} & -\frac{89}{1-\sigma^2} & \frac{1-\sigma^2}{4} 
    \end{bmatrix}
\end{equation}
and its determinant is
\begin{equation}
    \det( I_3 - J_{\alpha}) = \frac{(1-\sigma^2)^2 \mu \alpha}{16} - \frac{ 356L^4\alpha^3}{\mu (1-\sigma^2)^2} - \frac{120\alpha^3 \mu L^2}{(1-\sigma^2)^2}.
\end{equation}
When  $0 < \alpha \leq \frac{(1-\sigma^2)^2}{187\Tilde{Q}L}$, $\det( I_3 - J_{\alpha}) \geq \frac{(1-\sigma^2)^2 \mu \alpha}{32} $, and

\begin{align*}
[\mathrm{adj}(I_3 - J_{\alpha})]_{1, 2} = \frac{178L^2\alpha^2}{(1-\sigma^2)^2}, \\ [\mathrm{adj}(I_3 - J_{\alpha})]_{1, 3} = \frac{\mu L^2\alpha^3}{(1-\sigma^2)}, \\
[\mathrm{adj}(I_3 - J_{\alpha})]_{2, 2} \leq \frac{(1-\sigma^2)^2}{8}, \\ [\mathrm{adj}(I_3 - J_{\alpha})]_{2, 3} = \frac{4 L^4\alpha^3}{\mu(1-\sigma^2)}, \\
[\mathrm{adj}(I_3 - J_{\alpha})]_{3, 2} = 44.5,\\ [\mathrm{adj}(I_3 - J_{\alpha})]_{3, 3} = \frac{\mu \alpha (1-\sigma^2)}{4}. \\
\end{align*}
Next, we derive a matrix upper-bounding (element-wise) $(I_3 - J_{\alpha})^{-1} H_{\alpha} = \frac{\mathrm{adj}(I_3 - J_{\alpha}) }{\det (I_3 - J_{\alpha}) }H_{\alpha}$.
For $0 \leq \alpha \leq \frac{(1-\sigma^2)^2}{187\Tilde{Q}L}$, 
\begin{equation}
    (I_3 - J_{\alpha})^{-1} H_{\alpha} \leq 
    \begin{bmatrix}
    0.039 & 0.039 & 0 \\
    0.23 & 0.23 & 0 \\
    \frac{335}{(1-\sigma^2)^2} & \frac{335}{(1-\sigma^2)^2} & 0
    \end{bmatrix}.
\end{equation}
If $\q = [1, 1, \frac{1457}{(1-\sigma^2)^2}]$, we have
\begin{equation}
    ((I_3 - J_{\alpha})^{-1}H_{\alpha})\q \leq 0.66\q.
\end{equation}
Finally, invoking the definition of the weighted matrix norm, we complete the proof of the second inequality in the lemma.
\end{proof}

This completes the presentation of auxiliary results that support the proof of Theorem~1 in the main paper.

\section{Experimental Results for the Decentralized Average Consensus Problem}

\begin{figure*}[!htbp]
	\centering
	\minipage[t]{1\linewidth}
	\begin{subfigure}[t]{.49\linewidth}
		\includegraphics[width=\textwidth]{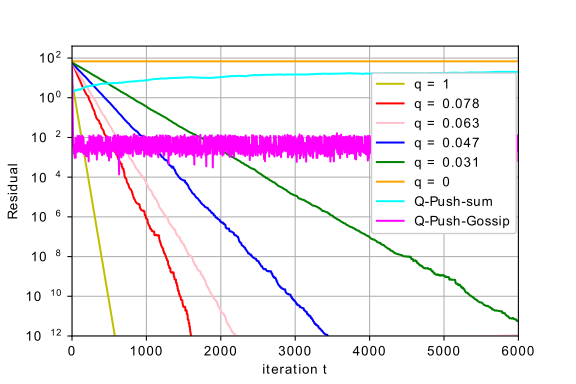}
		\caption{\footnotesize  Consensus residual: $\mathcal{B} = 1$}
	\end{subfigure}
	\begin{subfigure}[t]{.49\linewidth}
		\includegraphics[width=\textwidth]{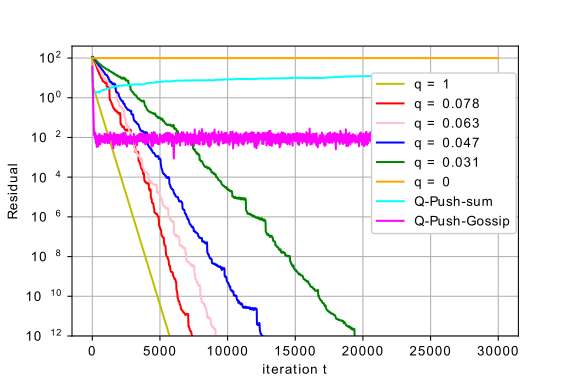}
		\caption{\footnotesize  Consensus residual: $\mathcal{B} = 10$}
	\end{subfigure}
	\caption{Average consensus on a jointly connected network with $\mathcal{B} = 1, 10, \epsilon = 0.05$.
	In each of the subplots, we show the performance of Di-CS-AC, i.e., Algorithm~1, for $6$ different sparsification levels and compare it to $2$ benchmark quantization algorithms, Q-Push-sum and Q-Push-Gossip. The quantization level is chosen such that the number of communicated bits for the benchmark algorithms is equal to that of Di-CS-AC when $q = 0.078$.
	} 
	\label{fig:Consensus_1}
	\endminipage 
		\vspace{-0.4cm}
\end{figure*}

In this section, we refer to the proposed Algorithm~1 in the main paper as Di-CS-AC (\underline{Di}rected \underline{C}ommunication-\underline{S}parsifying \underline{A}verage \underline{C}onsensus) and present its performance. 

We consider an average consensus problem where the dimension of a local parameter vector at each node is $d=64$. The initial state $\mathbf{x}_i^0$ is randomly generated from the normal distribution; the goal of the network is to reach the average consensus vector, i.e., compute $\Bar{\mathbf{x}}=\frac{1}{n}\sum_{i=1}^n \mathbf{x}_i^0$. The network setups are exactly the same as the decentralized optimization experiments in Section V.

For benchmarking purposes, we consider two quantized versions of the push-sum algorithm: 
(i) Q-Push-sum, obtained by applying simple quantization to the push-sum scheme \cite{kempe2003gossip,nedic2014distributed}, and (ii) Q-Push-Gossip, a quantized push-sum for gossip algorithm recently proposed in \cite{taheriquantized}. The former was originally developed for unconstrainted communication settings, while the latter originally targeted static networks; in the absence of prior work on communication-constrained consensus over time-varying directed networks, we adopt these two as the benchmarking schemes.

We compare the performance of different algorithms by computing the residual value $\frac{\|\mathbf{x}^t-\Bar{\mathbf{x}}\|}{\|\mathbf{x}^0-\Bar{\mathbf{x}}\|}$;
the results are shown in Fig. \ref{fig:Consensus_1}. As the figures demonstrate, at the considered levels of sparsification $q$ and values of the connectivity parameter $\mathcal{B}$, Di-CS-AC converges to the same limit as the full communication schemes. The convergence rate is linear in the number of iterations $t$ but smaller compression level and larger connectivity period slow the convergence down.
In Fig. \ref{fig:Consensus_1} (a) and (b), the two benchmarking quantization algorithms cannot reach the desired consensus accuracy in the time-varying directed network while the proposed Di-CS-AC achieves considerably smaller consensus error.

In regards to communication cost, to reach the same residual threshold (i.e., $10^{-12}$), our consensus algorithm, Di-CS-AC, with $q = 0.078$, incurs only $7.8 \%$ per-iteration cost as compared to the non-compression consensus algorithm (with $q = 1$) and needs fewer than double the number of iterations (see Fig. \ref{fig:Consensus_1} (a) and (b)). This demonstrates communication efficiency of Di-CS-AC.

\section{Experimental results for larger size of network}
\begin{figure}[h]
    \centering
    \includegraphics{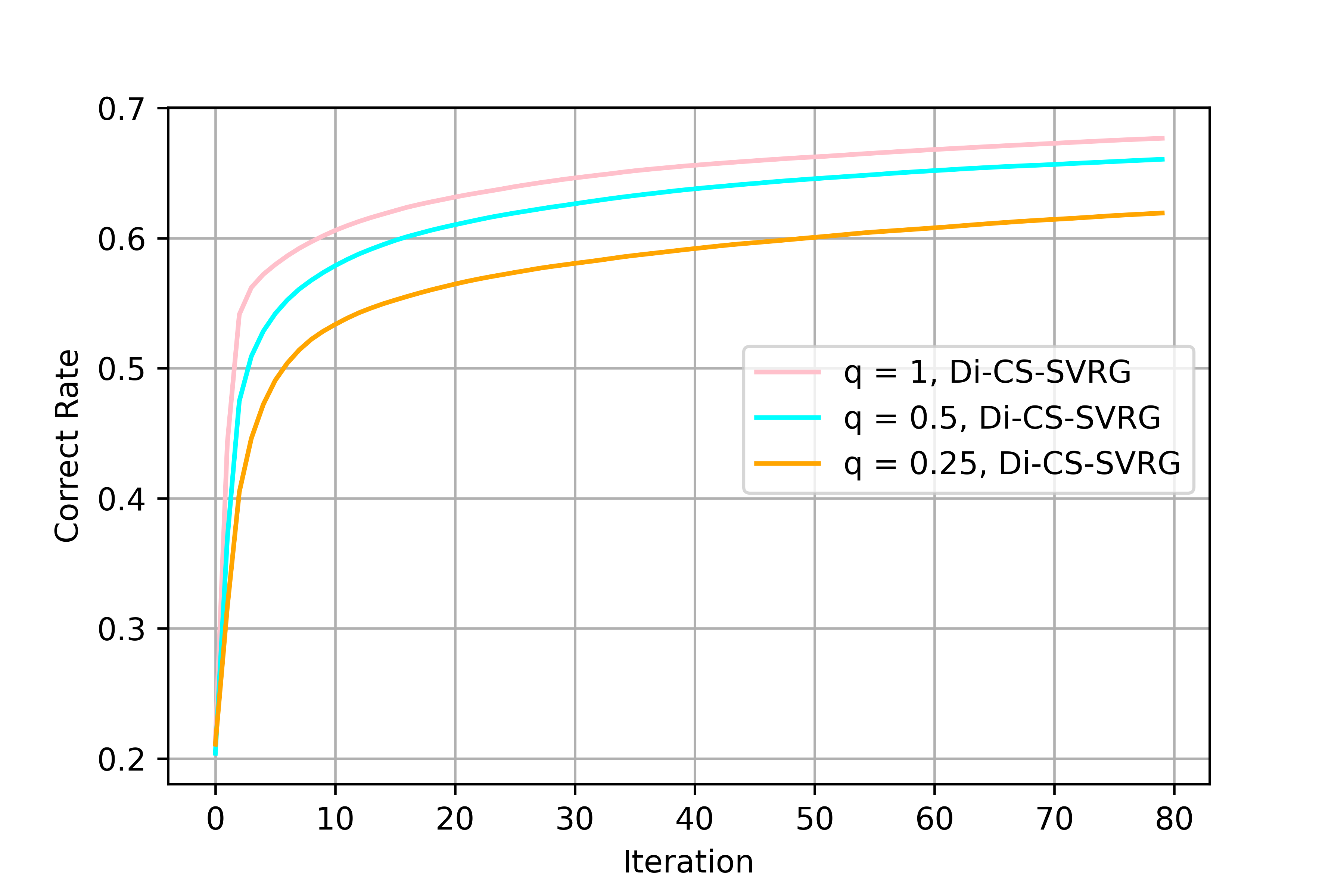}
    \caption{Di-CS-SVRG performance on a time-varying directed network with $100$ nodes.  }
    \label{fig:100nodes}
\end{figure}

In this section, we consider a network with $100$ nodes and repeat the logistic regression experiment. We still use the stackoverflow dataset introduced in Section~V of the main paper. The network is constructed following the same steps introduced in Section~V of the main paper with parameter $B = 1$. In Fig. \ref{fig:100nodes}, we show the correct rate for three different sparsification levels, i.e., $q = 0.25, 0.5 $ and $1$.

As the plot shows, schemes with different sparsification levels have different convergence rates -- more aggressive sparsification leads to slower convergence.

\end{document}